\documentclass[]{fairmeta}

\usepackage{amsmath}
\usepackage{tabularx}
\usepackage{enumitem}
\usepackage{xspace}
\usepackage[table]{xcolor}

\usepackage{booktabs}
\usepackage{pifont}
\usepackage{wasysym}
\usepackage{colortbl} 
\usepackage{wrapfig}
\usepackage{multirow}

\newcommand{\cmark}{\ding{51}}
\newcommand{\xmark}{\ding{55}}

\newcommand{\vs}{\textit{vs.}\xspace}
\setlist[itemize]{leftmargin=*, topsep=2pt, itemsep=1pt}
\setlist[enumerate]{leftmargin=*, topsep=2pt, itemsep=1pt}

\crefname{section}{Sec.}{Secs.}
\Crefname{section}{Section}{Sections}
\Crefname{table}{Table}{Tables}
\crefname{table}{Tab.}{Tabs.}

\newtcolorbox{instructionbox}[1][]{
  colback=cyan!2,
  colframe=cyan!35!black,
  boxrule=0.5pt,
  arc=2pt,
  left=7pt,
  right=7pt,
  top=6pt,
  bottom=6pt,
  before skip=10pt,
  after skip=10pt,
  fonttitle=\bfseries,
  title={Task\if\relax\detokenize{#1}\relax\else \; \textperiodcentered \; {\footnotesize #1}\fi},
  breakable
}

\newcommand{\benchname}{\mbox{TUA-Bench}\xspace}

\title{\raisebox{-0.16\height}{\includegraphics[height=1.1em]{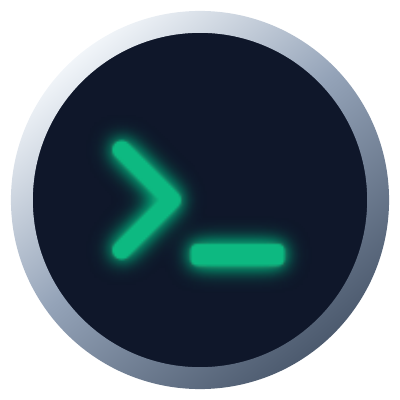}}\hspace{0.2em} TUA-Bench: A Benchmark for General-Purpose Terminal-Use Agents}

\author[1,*]{\fontsize{9}{11}\selectfont Shoufa Chen}
\author[1,*]{\fontsize{9}{11}\selectfont Luyuan Wang}
\author[2]{\fontsize{9}{11}\selectfont Xuan Yang}
\author[1]{\fontsize{9}{11}\selectfont Zhiheng Liu}
\author[1]{\fontsize{9}{11}\selectfont Yuren Cong}
\author[3]{\fontsize{9}{11}\selectfont Yuanfeng Ji}
\author[1]{\fontsize{9}{11}\selectfont Feiyan Zhou}
\author[1]{\fontsize{9}{11}\selectfont Xiaohui Zhang}
\author[1]{\fontsize{9}{11}\selectfont Fanny Yang}
\author[1]{\fontsize{9}{11}\selectfont Belinda Zeng}

\affiliation[1]{Meta AI}
\affiliation[2]{Duke University}
\affiliation[3]{Stanford University}
\contribution[*]{Equal contribution}

\abstract{As large language models and harness frameworks continue to advance, agents operating in terminals are increasingly capable of performing a broader range of general computer-use tasks beyond coding. However, existing benchmarks do not adequately evaluate general-purpose terminal computer-use agents (TUAs): general computer-use benchmarks primarily target graphical user interfaces (GUIs), whereas terminal-based benchmarks largely emphasize technical and programming-centric workflows historically native to the shell. 
We introduce TUA-Bench, a general-purpose benchmark for terminal-use agents. TUA-Bench includes 120 real-world tasks across five task families, covering routine digital activities—including document editing, email management, and live-web information seeking—as well as scientific and engineering workflows co-designed with PhD-level domain experts that require specialized software. This breadth distinguishes \benchname from prior shell-focused or domain-specific benchmarks. Each task is manually designed, runs in a real terminal with a deterministic setup script, and is evaluated by an execution-based scoring protocol. We find that the strongest frontier agent, Claude Code with Claude Opus 4.8 \texttt{max} reasoning effort, achieves 65.8\% overall performance, with substantial gaps across both tracks. By providing a broad and realistic evaluation of terminal-use capabilities, TUA-Bench aims to accelerate the transition from narrow, task-specific assistants to general-purpose agents capable of operating reliably across diverse digital environments.
}

\date{\today}
\metadata[Website]{\url{https://tuabench.ai/}}
\metadata[Code]{\url{https://github.com/facebookresearch/TUA-Bench}}

\begin{document}

\maketitle

\begin{figure}
    \centering
    \small
    \includegraphics[width=0.96\linewidth]{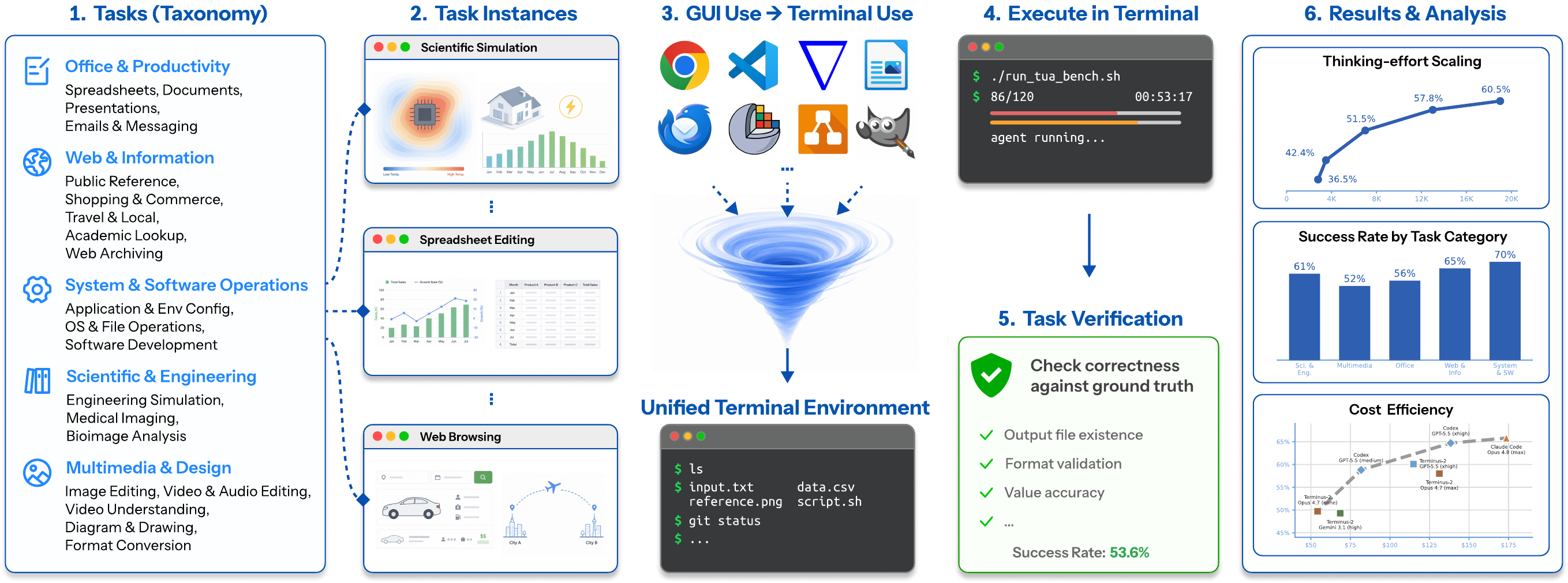}
   \caption{\textbf{Overview of TUA-Bench.} TUA-Bench evaluates terminal-use agents on realistic, application-grounded tasks spanning a five-domain taxonomy of real-world workflows. Each workflow is instantiated as concrete tasks in a unified terminal environment. Tasks that would conventionally require graphical interfaces are reformulated as GUI-to-terminal problems, requiring agents to interact solely through the command line. Agents execute each task autonomously, and the resulting rollout is automatically verified against ground truth.}\label{fig:teaser}
   \vspace{-40pt}
\end{figure}

\section{Introduction}\label{sec:intro}

Large language models (LLMs) have catalyzed a paradigm shift in artificial intelligence, expanding from conversational tools~\citep{openai2022chatgpt, anthropic2023claude} to programming assistants~\citep{github2021copilot, anysphere2026cursor, openai2025codex, anthropic2026claudecode, opencode2026github, google2025geminicli} and, more recently, to autonomous agents capable of executing complex, multi-step workflows beyond programming~\citep{manus2025launch, openclaw2026}. As these agents increasingly act on behalf of users across diverse digital environments, evaluating their ability to use computers reliably has become an important problem.

Most existing computer-use agents and benchmarks~\citep{zhou2024webarena, xie2024osworld, jia2026osworldmcp} assume that agents interact with computers through graphical user interfaces~(GUIs), such as desktop applications or web-based interfaces, mirroring how humans operate software systems. While GUIs are natural for human users, they are not necessarily the most suitable interface for LLM-based agents. GUI operation requires agents to combine language reasoning with visual perception: they must interpret screenshots, ground actions to precise screen locations, and remain robust to changes in layout, resolution, and rendering. These requirements introduce perception and grounding challenges, causing GUI-based evaluations to partly measure visual understanding and coordinate-level control rather than an agent's core ability to plan, reason, and use tools.

Command-line interfaces (CLIs), in contrast, expose computer interaction in a text-native form. Commands are explicit, feedback is textual, and complex workflows can be composed through scripts, pipes, and specialized programs. These properties make CLI environments naturally aligned with the strengths of language models. Moreover, many high-value professional workflows, including software engineering, data analysis, scientific computing, system administration, and multimedia processing, are already conducted primarily through terminals~\citep{kernighan1979unix,janssens2014data,piccolo2016tools}.

Beyond these traditionally terminal-centric domains, command-line access is increasingly available for a broader range of software systems. Popular platforms provide official or widely used command-line tools, including GitHub~\citep{githubcli}, Slack~\citep{slackcli}, Google Cloud~\citep{gcloudcli}, and Lark~\citep{larksuite_cli}; community-developed projects such as OpenCLI~\citep{opencli2026} and CLI-Anything~\citep{yang2026cli} further extend command-line access across application ecosystems. Together, these developments suggest that terminal interfaces are evolving from programming-specific tools into a broader medium for invoking applications, inspecting state, composing tools, and verifying results.

\begin{wrapfigure}{r}{0.4\linewidth}
    \centering
    \small
    \vspace{-5pt}
    \includegraphics[width=\linewidth]{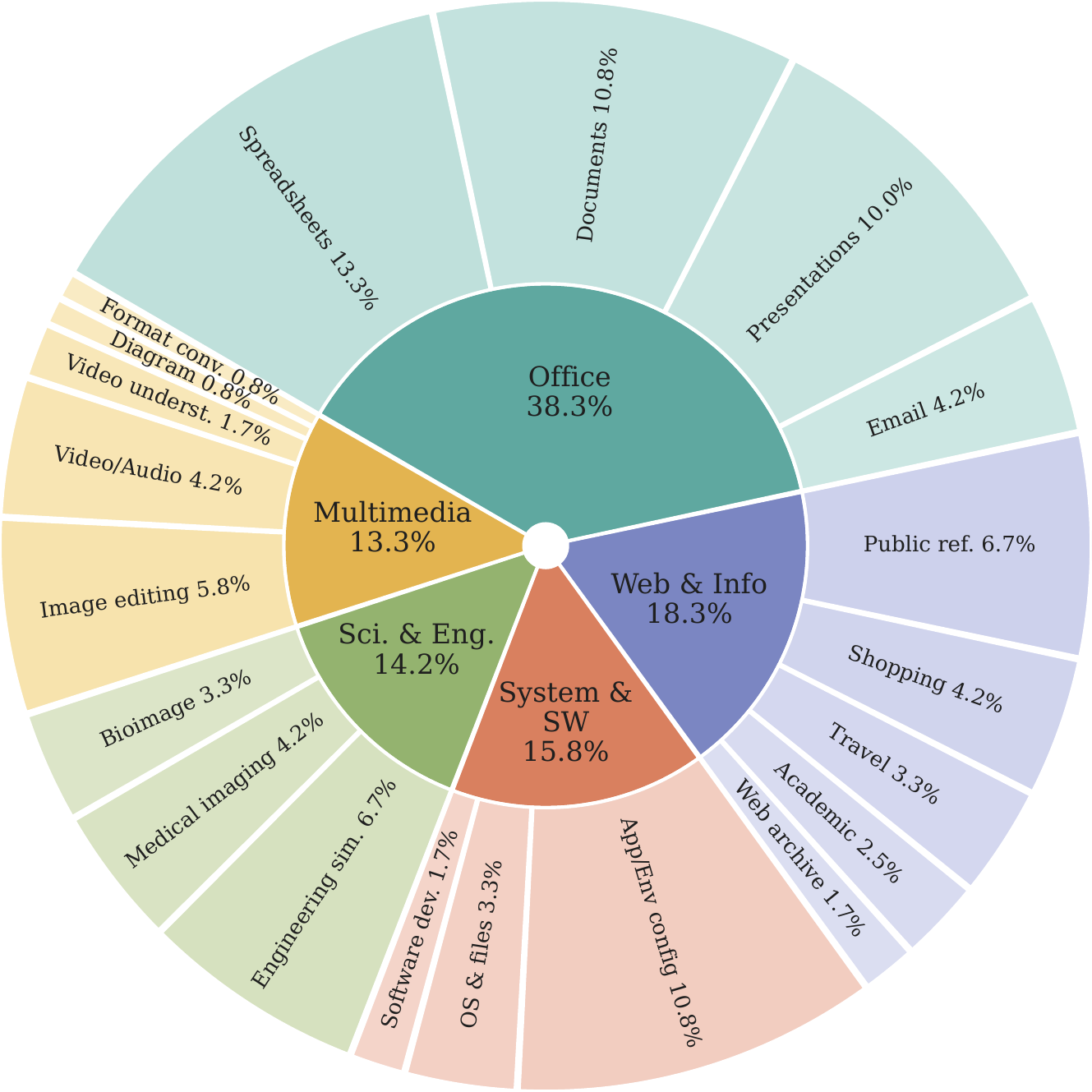}
    \vspace{2pt}
    \caption{\textbf{TUA-Bench task distribution.} The 120 tasks span five categories with fine-grained subcategories, covering both everyday digital work and expert professional workflows.}\label{fig:task-distribution}
    \vspace{-12pt}
\end{wrapfigure}
Despite this trend, existing evaluations have not fully captured the breadth of terminal-based computer use. Current terminal benchmarks, such as Terminal-Bench~\citep{merrill2026terminalbench},
primarily focus on shell-native technical and programming workflows, and therefore remain limited in evaluating general-purpose terminal-based computer use.

Motivated by this gap, we introduce \benchname, a benchmark for evaluating general-purpose terminal-use agents. As illustrated in \Cref{fig:teaser}, \benchname contains 120 diverse, realistic, and challenging computer-use tasks that require agents to operate through CLI, organized into five task families that span the breadth of computer-use, from everyday digital work such as document editing, web information seeking, and media processing, to expert scientific and engineering workflows co-designed with PhD-level domain experts in biology, medical physics, architectural engineering, and mechanical engineering. To ensure task quality and difficulty, we start from an initial pool of candidate tasks and apply a rigorous curation process, filtering or revising tasks with ambiguous instructions, overly simple solutions, or mismatches between input files and target outputs. The final benchmark is manually verified and designed to provide a reliable, challenging, and informative evaluation of terminal-use agents.

We further conduct a comprehensive evaluation of frontier models and agent frameworks on \benchname, including ablations on model reasoning effort. The strongest evaluated agent achieves 65.8\% success rate, revealing substantial remaining gaps in long-horizon planning, tool use, execution monitoring, and error recovery in terminal environments. By providing a broad and realistic testbed for terminal-based computer use, \benchname aims to support progress toward general-purpose agents that can operate reliably across diverse digital and professional workflows.

Our main contributions are summarized as follows:

\begin{itemize}
    \item We introduce \benchname, a high-quality benchmark for evaluating general-purpose terminal-use agents. It contains 120 diverse, realistic, and challenging tasks that span everyday digital work and professional workflows co-designed with domain experts.

    \item We provide a reliable evaluation suite for benchmarking terminal-use agents. Each task is specified in a standardized format, paired with executable environments and verification procedures, and manually curated from an initial pool of 394 candidates to remove ambiguous, overly simple, or inconsistent tasks.

    \item We conduct a comprehensive evaluation of frontier models and agent frameworks on \benchname, including ablations on model reasoning effort. Our analysis reveals that even the strongest evaluated agent achieves only 65.8\% success rate, highlighting remaining challenges in long-horizon planning, tool use, execution monitoring, and error recovery.
\end{itemize}


\section{Related Work}
\label{sec:related-work}

\noindent\textbf{GUI-based computer-use benchmarks.}
Most benchmarks for general computer use evaluate agents through GUI-based interaction, reflecting the central role of graphical interfaces in human-computer interaction. This line of work spans controlled browser tasks in MiniWoB~\citep{shi2017world} and MiniWoB++~\citep{zheran2018reinforcement}, real-world and reproducible web environments in Mind2Web~\citep{deng2023mind2web} and WebArena~\citep{zhou2024webarena}, live-web evaluation in WebVoyager~\citep{he2024webvoyager}, and full desktop operation in OSWorld~\citep{xie2024osworld}, WindowsAgentArena~\citep{bonatti2025windows}, macOSWorld~\citep{yang2026macosworld}, and OSUniverse~\citep{davydova2025osuniverse}. These benchmarks evaluate both general agentic capabilities, such as planning and error recovery, and GUI-specific skills, such as visual grounding, spatial reasoning, and graphical control. They therefore provide coverage of graphical computer use, while leaving open how well agents can operate computers through text-native command interfaces.

\noindent\textbf{CLI-based agents and benchmarks.}
Recent agent systems increasingly treat the command line as a practical interface for computer operation. SWE-agent~\citep{yang2024swe} and OpenHands~\citep{wang2025openhands} demonstrate how agents can use shell access to inspect codebases, edit files, execute tests, and iteratively resolve software tasks. This trend has also appeared in deployed CLI assistants, including Claude Code~\citep{anthropic2026claudecode}, Codex CLI~\citep{openai2025codex}, Gemini CLI~\citep{google2025geminicli}, OpenCode~\citep{opencode2026github}, Qwen Code~\citep{qwen_code2026}, Kimi Code~\citep{kimi_code2026}, and MiMo Code~\citep{mimo_code2026}. These systems highlight the growing importance of CLI-based interaction, although their primary focus remains software engineering and developer assistance.

The emergence of CLI-based agents has motivated benchmarks for text-native interaction, where agents issue commands, invoke tools, and produce verifiable outcomes rather than manipulate pixels. Closest to our setting, Terminal-Bench~\citep{merrill2026terminalbench} evaluates long-horizon terminal tasks with executable tests, while TerminalWorld~\citep{chu2026terminalworld} scales this paradigm using real-world terminal recordings. Other benchmarks target specialized command-line capabilities, including shell programming and optimization in Koala~\citep{lamprou2025koala}, development-environment configuration in SetupBench~\citep{arora2025setupbench} and the process-level framework of \citet{kuang2025process}, natural-language-to-Bash translation in \citet{westenfelder2025llm}, and reward-hacking vulnerabilities in Terminal-Wrench~\citep{bercovich2026terminal}. SWE-bench Verified~\citep{jimenez2023swe} is also execution-grounded but focuses on repository-level software repair. Related tool-use benchmarks such as MCP-Universe~\citep{luo2025mcp} and Toolathlon~\citep{li2026the} evaluate agents through structured tool interfaces. Overall, existing CLI-based and text-native evaluations remain concentrated on software engineering, technical terminal workflows, or structured tool use.

\noindent\textbf{Positioning of \benchname.}
As summarized in \Cref{tab:positioning}, existing benchmarks can be situated along two central dimensions: the interface through which agents act and the scope of work they evaluate. Benchmarks with broad or professionally relevant task coverage generally rely on graphical or application-specific interfaces. OSWorld~\citep{xie2024osworld} evaluates general desktop interaction through a GUI, while OfficeBench~\citep{wang2024officebench} focuses on workflows across office applications. ScienceAgentBench~\citep{chen2025scienceagentbench} and GDPval~\citep{patwardhan2025gdpval} extend evaluation to economically valuable tasks. By contrast, existing terminal benchmarks remain concentrated on technical and shell-centric workflows, leaving broad computer use through the terminal largely unevaluated. \benchname fills this gap by combining native command-line interaction with broad task coverage. It evaluates everyday, technical, and expert work in executable terminal environments, providing a unified, execution-grounded assessment of whether agents can use the terminal as a general-purpose interface to computers.

\begin{table*}[t]
\small
\centering
\caption{\textbf{Positioning of \benchname.} We compare \benchname with representative agent benchmarks along interface modality, multimodal requirements, task scope, scientific coverage, and scale. \textbf{Interface} denotes the modality through which the agent acts. \textbf{MM} indicates whether tasks require perception or processing of non-textual inputs. \textbf{Task scope} summarizes coverage of office productivity, web use, and software/system operations. \textbf{\# Sci. Subj.} denotes the number of scientific subjects covered, counted according to the discipline-level taxonomy of ScienceAgentBench~\citep{chen2025scienceagentbench}. \cmark~/~\xmark~indicate covered/not covered, and \textbf{\# Tasks} denotes the total number of tasks.}\label{tab:positioning}
\setlength{\tabcolsep}{4pt}
\renewcommand{\arraystretch}{1.15}
\begin{tabular}{l l c c c c c c}
\toprule
\multirow{2}{*}{Benchmark} & \multirow{2}{*}{Interface} & \multirow{2}{*}{MM} & \multicolumn{3}{c}{Task scope} & \multirow{2}{*}{\# Sci.\ Subj.} & \multirow{2}{*}{\# Tasks} \\
\cmidrule(lr){4-6} & & & Office & Web & SWE/Sys. & & \\
\midrule
WebArena~\citep{zhou2024webarena}
& Web GUI & \xmark & \xmark & \cmark & \xmark & 0 & 812 \\
 
OSWorld~\citep{xie2024osworld}
& Desktop GUI & \cmark & \cmark & \cmark & \cmark & 0 & 369 \\
 
MCP-Universe~\citep{luo2025mcp}
& MCP tools & \xmark & \xmark & \cmark & \cmark & 0 & 231 \\
 
SWE-bench Verified~\citep{jimenez2023swe}
& Code repo & \xmark & \xmark & \xmark & \cmark & 0 & 500 \\
 
ScienceAgentBench~\citep{chen2025scienceagentbench}
& Code scaffold & \xmark & \xmark & \xmark & \xmark & 4 & 102 \\
 
Terminal-Bench 2.0~\citep{merrill2026terminalbench}
& Terminal & \xmark & \xmark & \xmark & \cmark & 0 & 89 \\
\midrule
\textbf{\benchname}~(ours)
& Terminal & \cmark & \cmark & \cmark & \cmark & 4 & 120 \\
\bottomrule
\end{tabular}
\end{table*}

\section{TUA-Bench}\label{sec:tua-bench}
TUA-Bench is a benchmark framework for evaluating the ability of agents and models to complete a broad range of terminal-based tasks, spanning both everyday digital workflows and domain-specialized professional procedures. The benchmark is designed to assess not only whether agents can execute isolated commands, but also whether they can plan, interact with realistic software environments, manipulate files and artifacts, and verify task completion under reproducible conditions. The following subsections first introduce the execution environment underlying TUA-Bench, and then describe the task curation process.

\subsection{Task Execution Environment}\label{sec:task-env}

To support standardized and reproducible evaluation, we build \benchname on top of Harbor~\citep{Harbor_Framework}, the orchestration framework also used by Terminal-Bench~\citep{merrill2026terminalbench}. Harbor provides a mature execution substrate for terminal-agent evaluation, including task setup, environment management, execution control, logging, and result verification. By adopting this shared infrastructure, \benchname can focus on constructing, curating, and validating realistic tasks while maintaining compatibility with existing terminal-agent evaluation pipelines.

\noindent\textbf{Execution infrastructure.}
Harbor serves as the orchestration layer of \benchname. It manages task configuration, environment construction, container launch, parallel execution, and the collection of trajectories, token usage, scores, and runtime metadata. Each task is executed inside an isolated and resettable Linux container, ensuring that evaluations are reproducible and that failed, incomplete, or unsafe executions do not affect subsequent trials. In addition to Docker, \benchname supports Podman~\citep{podman2026}, which preserves Dockerfile compatibility while enabling rootless execution on shared clusters without requiring \texttt{sudo} privileges. This infrastructure allows agents to operate in realistic terminal environments, interacting with files, shell commands, installed packages, optional internet access, and native CLI-based agent interfaces.

\noindent\textbf{Task specification and reproducibility.}
Each task in \benchname is packaged as a self-contained specification, including a Dockerfile, task-specific input artifacts, natural-language instructions, environment variables, model and runtime settings, and an in-environment verifier. This packaging standardizes the initial state, execution procedure, and evaluation protocol for every task. As a result, \benchname provides consistent and reproducible evaluation while preserving realistic sources of variability, such as stochastic agent behavior and internet-dependent execution when network access is enabled.

\subsection{Task Curation}\label{sec:task-curation}
TUA-Bench comprises 120 real-world tasks organized along two complementary dimensions. The \emph{breadth} dimension captures the everyday digital work of general computer users, including web browsing, document and spreadsheet editing, email management, and media processing. The \emph{depth} dimension focuses on expert workflows co-designed with PhD-level domain experts in biology, medical physics, architectural engineering, and mechanical engineering. These tasks require agents to operate specialized software and complete domain-specific procedures, often through command-line interfaces.

Our curation pipeline reflects the different design requirements of everyday and professional computer-use tasks. For everyday tasks~(\cref{sec:everyday-task}), we start from established GUI-based benchmarks and translate them into terminal-centric settings, preserving the underlying user goals while requiring agents to complete the tasks through command-line interaction and file-level manipulation. For professional tasks~(\cref{sec:pro-tasks}), we construct new workflows in collaboration with domain experts, with an emphasis on realistic procedures, domain-specific constraints, and executable evaluation. We describe these two tracks in detail below.

\subsubsection{Everyday Digital Tasks}\label{sec:everyday-task}
We source everyday digital tasks from OSWorld~\citep{xie2024osworld}, a GUI-based benchmark for evaluating computer-use agents on realistic tasks across real-world applications.

\noindent\textbf{GUI-to-CLI task conversion.}
OSWorld contains 369 computer-use tasks grounded in real web and desktop applications across open-ended domains. We reuse these tasks as the basis for our everyday digital task set and convert them into CLI-based tasks that can be executed within our evaluation environment~(\cref{sec:task-env}). For each task, we also reuse the corresponding input files and gold artifacts from OSWorld, which define the initial task state and the target output for evaluation. Unlike OSWorld, which specifies the target application during task setup, such as using LibreOffice to edit a spreadsheet, our formulation preserves only the underlying task intent and does not constrain the tools that an agent may use. This design gives agents greater flexibility in choosing appropriate commands, utilities, or workflows to complete each task. Based on this principle, we rewrite the task instructions so that they are natural and actionable in a terminal-based setting.

\noindent\textbf{Quality control via human verification.}
After converting all 369 tasks from OSWorld, we apply a rigorous quality-control process with human verification to identify and remove tasks with inconsistencies between the provided input files and the target gold artifacts.
Specifically, we manually inspect failed execution trajectories together with the final artifacts produced by agents. This human verification process reveals that, in some cases, task failure is caused not by agent limitations but by discrepancies between the task inputs and the corresponding gold files. For instance, certain presentation-editing tasks exhibit mismatches in slide themes or formatting conventions between the input and gold files. In these cases, an agent may correctly follow the task intent yet still fail the verifier due to artifact-level inconsistencies, making the task unreliable for evaluation. We therefore exclude tasks with verified input--gold mismatches from the benchmark.

\noindent\textbf{Difficulty-aware task selection.}
Benchmarks for agentic computer use can rapidly become saturated as foundation models and agent frameworks continue to improve. For instance, while the best-performing model achieved only a 12.24\% success rate on OSWorld at the time of its release, recent state-of-the-art GPT-5.5~\citep{openai2026introducinggpt55} reports substantially higher performance, reaching 78.7\%. This rapid progress suggests that many tasks that once served as meaningful challenges may no longer sufficiently distinguish the capabilities of modern agents.

\begin{table}[t]
\centering
\small
\caption{Scientific subjects spanning biology, medical physics, architectural engineering, and mechanical engineering. Each row shows the subject's task scope and a representative task example.}\label{tab:depth-subjects}
\renewcommand{\arraystretch}{1.1}
\begin{tabular}{@{}m{3.2cm}m{7.4cm}>{\centering\arraybackslash}m{4.4cm}@{}}
\toprule
\textbf{Subject} & \textbf{Scope} & \textbf{Task Example} \\
\midrule
Biology
& Counting and localizing cell nuclei from fluorescence / nuclear-stain micrographs; image-based cytometry of cells and subcellular structures.
& \includegraphics[width=\linewidth,height=2.2cm,keepaspectratio]{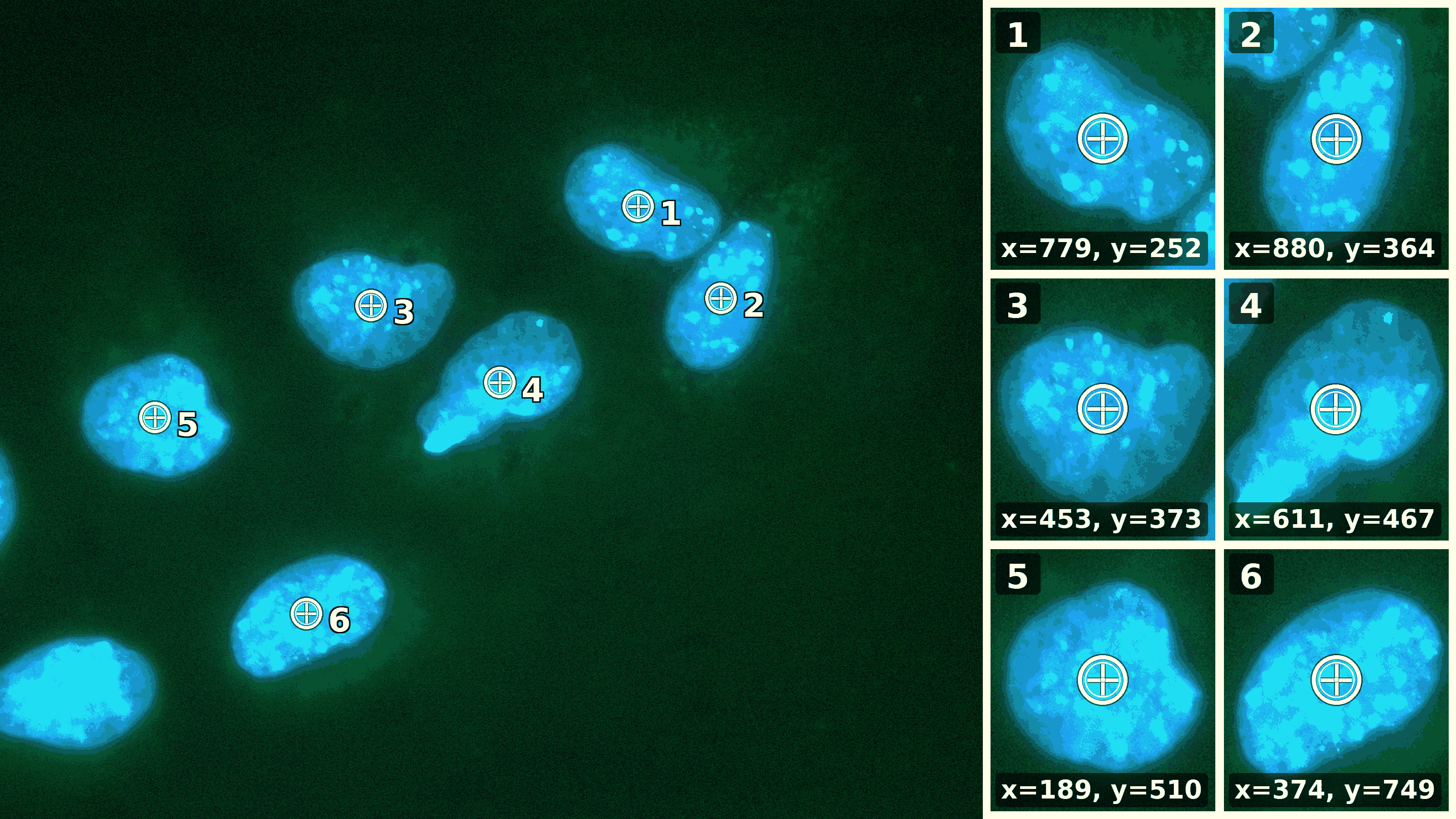}  \\
\addlinespace
Medical Physics
& Histopathology image segmentation, volumetry, and overlay from MRI volumes; anatomical segmentation and morphometry in medical image computing.
& \includegraphics[width=\linewidth,height=2.2cm,keepaspectratio]{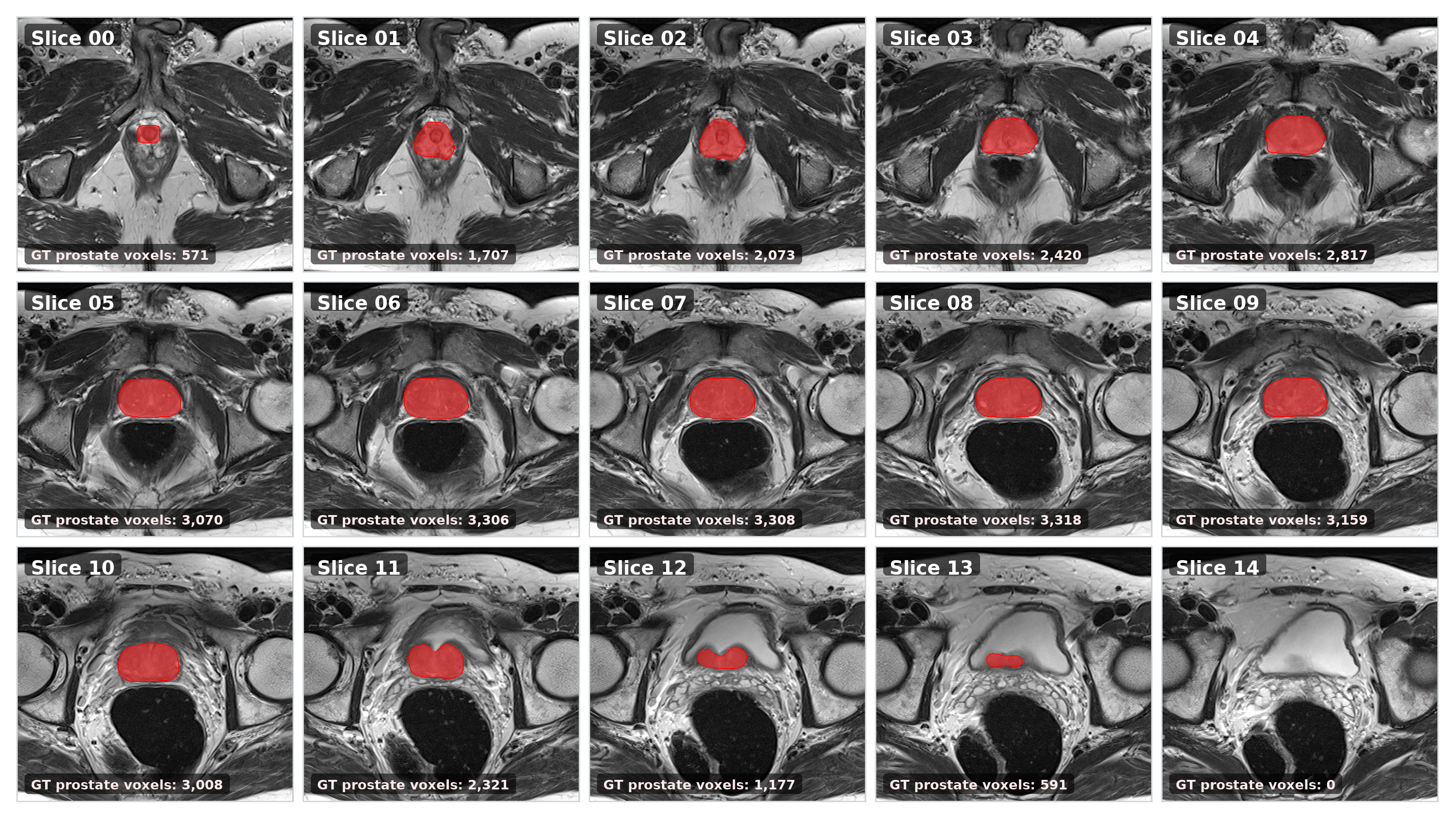} \\
\addlinespace
Architectural Engineering
& Reconstructing and simulating whole-building energy performance with OpenStudio and EnergyPlus; building energy performance simulation.
& \includegraphics[width=\linewidth,height=2.2cm,keepaspectratio]{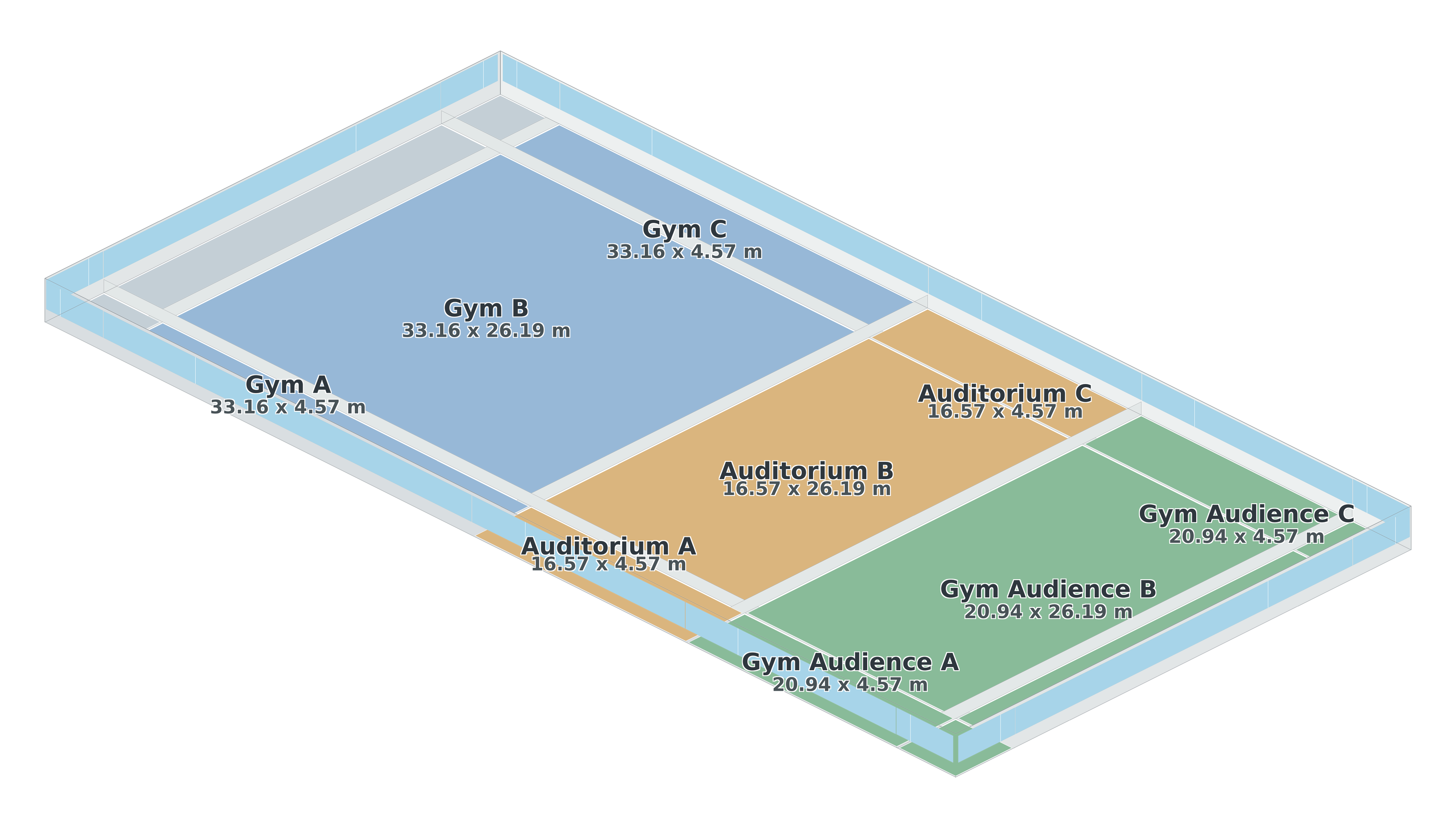} \\
\addlinespace
Mechanical Engineering
& Heater placement, cold-plate optimization, and heated-plate analysis via OpenFOAM, including conjugate heat transfer; computational fluid dynamics and heat transfer.
& \includegraphics[width=\linewidth,height=2.2cm,keepaspectratio]{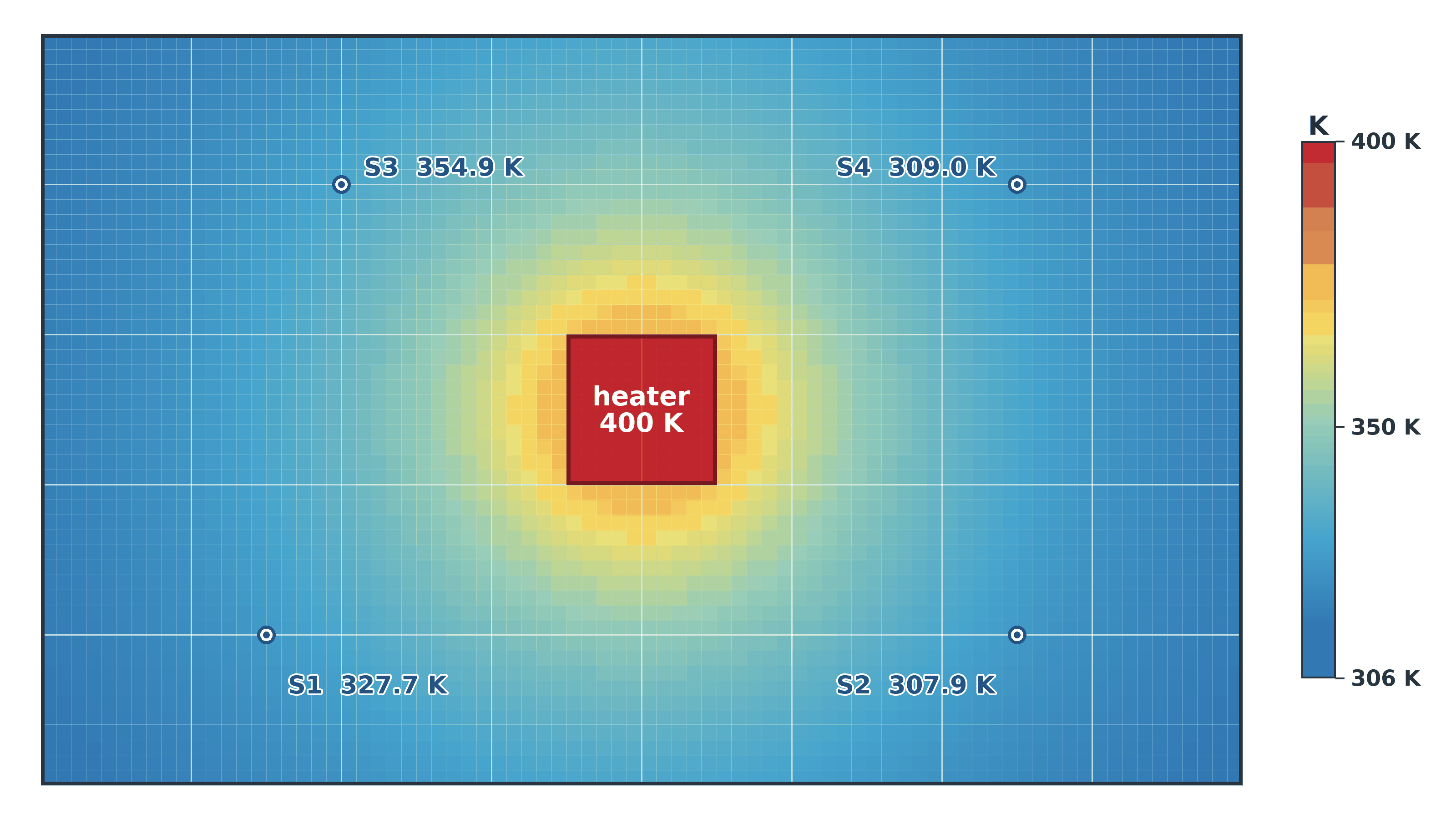} \\
\bottomrule
\end{tabular}
\end{table}

To this end, we evaluate each candidate task using three frontier models, GPT-5.5~\citep{openai2026introducinggpt55}, Claude Opus 4.7~\citep{anthropic2026opus47}, and Gemini 3.1 Pro~\citep{google2026gemini31pro}, each deployed within the Terminus-2 agent framework. For each model-task pair, we conduct five independent trials and compute the mean reward as a quantitative measure of empirical task solvability. Tasks are then ranked according to their aggregate solvability across models, and 100 tasks with the lowest solvability are retained for inclusion in the benchmark. Through this difficulty-aware selection process, we aim to construct a benchmark with sustained difficulty and long-term discriminative power by selecting tasks that remain among the most challenging under strong contemporary baselines.

\subsubsection{Professional Scientific Tasks}\label{sec:pro-tasks}
Beyond everyday digital tasks, we introduce a set of professional tasks designed to evaluate whether LLM agents can operate specialized software in domain-specific settings. These tasks target scientific and engineering workflows that require procedural knowledge, familiarity with specialized tools, and the ability to produce outputs satisfying domain-specific constraints. To ensure realism and evaluability, our design follows three principles. First, each task should reflect a meaningful workflow commonly performed by professionals, rather than an artificial exercise created solely for benchmarking. Second, each task should require a sufficiently complex sequence of operations, allowing us to assess whether agents can plan, execute, and revise multi-step procedures across multiple tool calls. Third, each task should support reliable evaluation, either through programmatic verifiers or LLM-as-a-judge assessment.

Concretely, we construct an initial pool of 25 professional tasks mainly across biology, medical physics, architectural engineering, and mechanical engineering. Each task is developed and validated in collaboration with PhD-level domain experts, who help identify representative workflows, prepare realistic input assets, specify expected outputs, and verify that the task reflects authentic professional practice. During task construction, we package the required files, software environments, and execution instructions into reproducible settings, so that agents must complete the workflow through actual interaction with specialized tools. We also define task-specific evaluation criteria, prioritizing automatically checkable outputs when possible and using expert-informed judging rubrics when programmatic verification is insufficient. Finally, we remove overly simple tasks that multiple agents can already solve and retain 20 challenging tasks.

\subsubsection{Task Statistics and Features}

\Cref{fig:task-distribution} presents a two-level taxonomy of the 120 tasks in TUA-Bench. The benchmark is structured into five top-level task families: Office \& Productivity, Web \& Information, System \& Software Operations, Scientific \& Engineering, and Multimedia \& Design. These families are further divided into 20 subcategories. In the sunburst visualization, the inner ring indicates the relative proportion of each top-level family, while the outer ring provides the corresponding subcategory-level breakdown. Office \& Productivity constitutes the largest portion of the benchmark, accounting for 38.3\% of all tasks, consistent with the prevalence of spreadsheet, document, presentation, and email workflows in everyday computer use. The remaining four families are more evenly represented, ranging from 13.3\% for Multimedia \& Design to 18.3\% for Web \& Information. This task composition is intended to cover both routine productivity scenarios and specialized professional workflows, including engineering simulation, medical imaging, and software configuration. As a result, aggregate performance on TUA-Bench reflects an agent's ability to operate across heterogeneous, real-world computer-use settings rather than within a narrow application domain.

\section{Benchmark Experiments}\label{sec:results}

\subsection{Experimental Settings}
\noindent\textbf{Agents.}
We conduct a broad evaluation of contemporary terminal-based agents and frontier language models to characterize both agent-level and model-level performance on \benchname. Specifically, we evaluate five agent frameworks: Terminus-2~\citep{harbor_terminus2}, Codex~\citep{openai2025codex}, OpenHands~\citep{wang2025openhands}, Mini-SWE-Agent~\citep{yang2024swe}, and Claude Code~\citep{anthropic2026claudecode}. 

\noindent\textbf{Models.}
To evaluate model capability across a wide range of current systems, we include both leading proprietary models and strong open or widely available alternatives. The evaluated models include GPT-5.5, GPT-5.4 mini, Claude Opus 4.8, Claude Opus 4.7, Claude Sonnet 4.6, Claude Haiku 4.5, Gemini 3.1 Pro, GLM-5.1, MiniMax-M3, DeepSeek-V4 Pro, Qwen3.7-Max, and Kimi K2.6. This model suite covers different capability tiers, ranging from frontier large-scale reasoning models to more compact or cost-efficient variants, enabling us to examine how \benchname distinguishes models across the performance spectrum.

\noindent\textbf{Metrics.}
We evaluate agents using execution-grounded task success rather than agent action/trajectory. Each task is associated with an automatic verifier that inspects the final environment state and returns a scalar reward, with full task completion corresponding to success. For each agent--model--thinking configuration, we run 5 independent trials per task and report the mean success rate across all trials. In addition, we report Pass@1, Pass@5, and All-5 to characterize reliability across repeated attempts: Pass@1 measures single-run task performance, Pass@5 measures the best outcome across five independent trials for each task, and All-5 measures the fraction of tasks solved consistently in all five trials. These complementary metrics distinguish systems that occasionally solve a task from those that solve it robustly.

\subsection{Main results}
\label{sec:main-results}

\begin{table}[t]
\centering
\small
\caption{\textbf{\benchname results.} \textbf{(a)} All models evaluated on the Terminus-2 scaffold, sorted by success rate. \textbf{(b)} Each agent at its best-performing model. All configurations use the \textbf{\emph{highest}} available reasoning-effort setting, except Claude Haiku 4.5$^\dagger$, for which thinking is disabled. \textbf{Bold} marks the best success rate within each panel. Additional agent-model combination results are reported in \Cref{tab:full-results}.}
\label{tab:two-panel}
\setlength{\tabcolsep}{10pt}
\renewcommand{\arraystretch}{1.05}
\begin{tabular}{l l c c c c}
\toprule
Agent & Model & Success Rate (\%) & Pass@1 & Pass@5 & All-5 \\
\midrule
\multicolumn{6}{l}{\textbf{(a)} Model sweep on the Terminus-2 agent} \\
\midrule
\multirow{12}{*}{Terminus-2}
 & GPT-5.5              & \textbf{60.1 $\pm$ 0.6} & 52.3\% & 64.2\% & 31.7\% \\
 & Claude Opus 4.8      & 59.7 $\pm$ 1.0 & 53.8\% & 62.5\% & 42.5\% \\
 & Claude Opus 4.7      & 58.0 $\pm$ 0.8 & 51.0\% & 64.2\% & 39.2\% \\
 & Gemini 3.1 Pro & 49.3 $\pm$ 1.8 & 41.2\% & 57.5\% & 24.2\% \\
 & GLM-5.1              & 48.1 $\pm$ 1.3 & 40.3\% & 59.2\% & 20.8\% \\
 & MiniMax-M3           & 47.0 $\pm$ 1.3 & 41.2\% & 59.2\% & 22.5\% \\
 & DeepSeek-V4 Pro      & 46.2 $\pm$ 0.8 & 38.0\% & 57.5\% & 18.3\% \\
 & Qwen3.7-Max          & 44.9 $\pm$ 0.7 & 37.7\% & 57.5\% & 21.7\% \\
 & Kimi K2.6            & 42.8 $\pm$ 1.8 & 35.3\% & 55.8\% & 18.3\% \\
 & Claude Sonnet 4.6    & 42.8 $\pm$ 0.3 & 34.8\% & 49.2\% & 20.0\% \\
 & GPT-5.4 mini         & 27.2 $\pm$ 1.4 & 20.0\% & 41.7\% & 6.7\%  \\
 & Claude Haiku 4.5$^\dagger$     & 23.9 $\pm$ 1.5 & 15.7\% & 30.8\% & 3.3\%  \\
\midrule
\multicolumn{6}{l}{\textbf{(b)} Best model per agent} \\
\midrule
Claude Code & Claude Opus 4.8 & \textbf{65.8 $\pm$ 0.7} & 58.8\% & 64.2\% & 51.7\% \\
Codex          & GPT-5.5          & 64.7 $\pm$ 0.7 & 57.7\% & 68.3\% & 42.5\% \\
OpenHands      & Claude Opus 4.8  & 63.4 $\pm$ 0.6 & 57.3\% & 67.5\% & 45.0\% \\
Mini-SWE-Agent & GPT-5.5          & 62.4 $\pm$ 0.8 & 54.2\% & 67.5\% & 40.0\% \\
Terminus-2     & GPT-5.5          & 60.1 $\pm$ 0.6 & 52.3\% & 64.2\% & 31.7\% \\
\bottomrule
\end{tabular}
\end{table}

\noindent\textbf{Model comparison at a fixed agent framework.}
\Cref{tab:two-panel}~(a) isolates the effect of the underlying model by fixing the agent to basic Terminus-2. Three frontier models form a leading group: GPT-5.5 (60.1\%), Claude Opus 4.8 (59.7\%), and Claude Opus 4.7 (58.0\%). The difference between the top two models is smaller than their trial-to-trial variation ($\pm0.6$ and $\pm1.0$, respectively), suggesting the observed difference is small relative to the reported run-to-run uncertainty. However, this near-tie in average performance does not imply comparable reliability. Claude Opus 4.8 solves all five attempts on 42.5\% of tasks, compared with 31.7\% for GPT-5.5, indicating more consistent behavior across seeds despite a similar overall success rate. Below this leading group, a gap of approximately nine percentage points separates the frontier models from a mid-tier band ranging from 44.9\% to 49.3\%, which includes Gemini 3.1 Pro Preview, GLM-5.1, MiniMax-M3, DeepSeek-V4 Pro, and Qwen3.7-Max. The within-family Claude results further reveal a clear capability hierarchy: Opus 4.8 achieves 59.7\%, Sonnet 4.6 achieves 42.8\%, and Haiku 4.5 achieves 23.9\%. This pattern suggests that the benchmark distinguishes model tiers effectively rather than being saturated.

\noindent\textbf{Best achievable performance per agent.}
\Cref{tab:two-panel}~(b) reports the strongest configuration for each agent scaffold by pairing every scaffold with its best-performing model. Claude Code achieves the highest overall result, reaching \textbf{65.8\%} with Claude Opus 4.8~(\texttt{max}), followed closely by Codex with GPT-5.5~(\texttt{xhigh}) at 64.7\%, OpenHands with Claude Opus 4.8~(\texttt{max}) at 63.4\%, and Mini-SWE-Agent with GPT-5.5~(\texttt{xhigh}) at 62.4\%. Terminus-2 with GPT-5.5~(\texttt{xhigh}) obtains 60.1\%, still within 5.7 percentage points of the best-performing configuration. Overall, the leading scaffolds occupy a relatively narrow performance band, suggesting that strong frontier models can yield competitive results across a range of agent implementations.

\subsection{Ablation Studies and Empirical Analysis}

\begin{figure}[t]
\centering
\begin{minipage}[t]{0.48\linewidth}
  \centering
  \includegraphics[width=\linewidth]{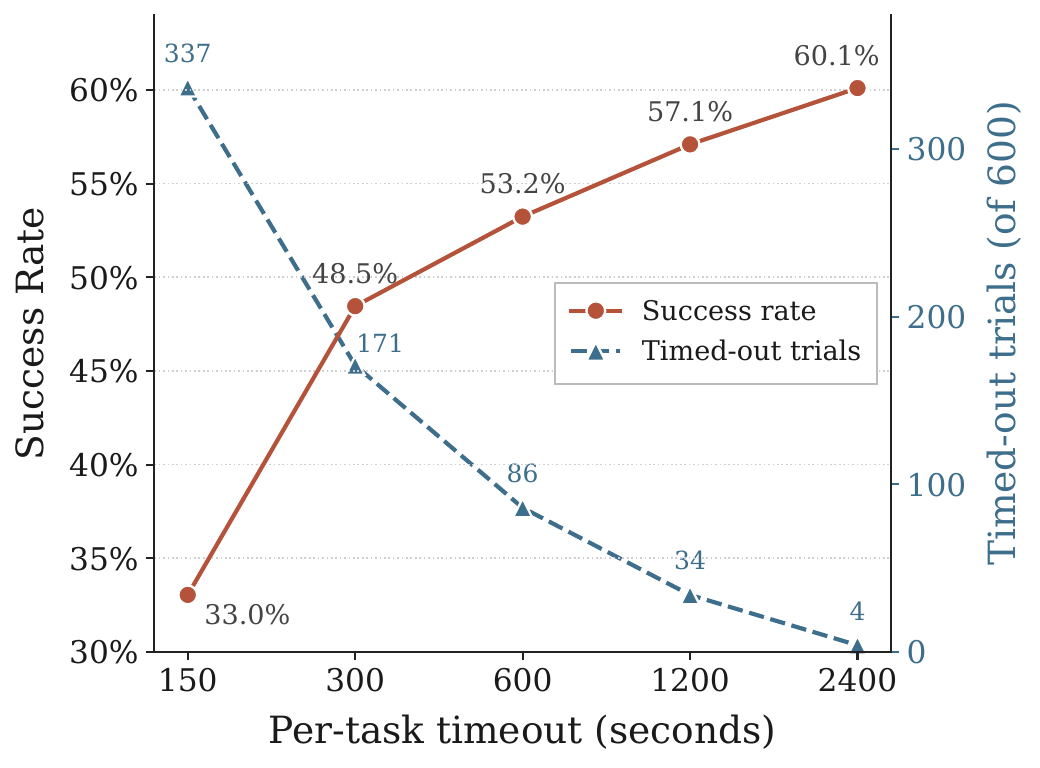}
  \caption{\textbf{Task-execution time budget} for Terminus-2 + GPT-5.5 (\texttt{xhigh}). Larger budgets cut timed-out trials from 337 to just 4 of 600 and steadily
raise the success rate.}\label{fig:time-budget-scaling}
\end{minipage}\hfill
\begin{minipage}[t]{0.48\linewidth}
  \centering
  \includegraphics[width=\linewidth]{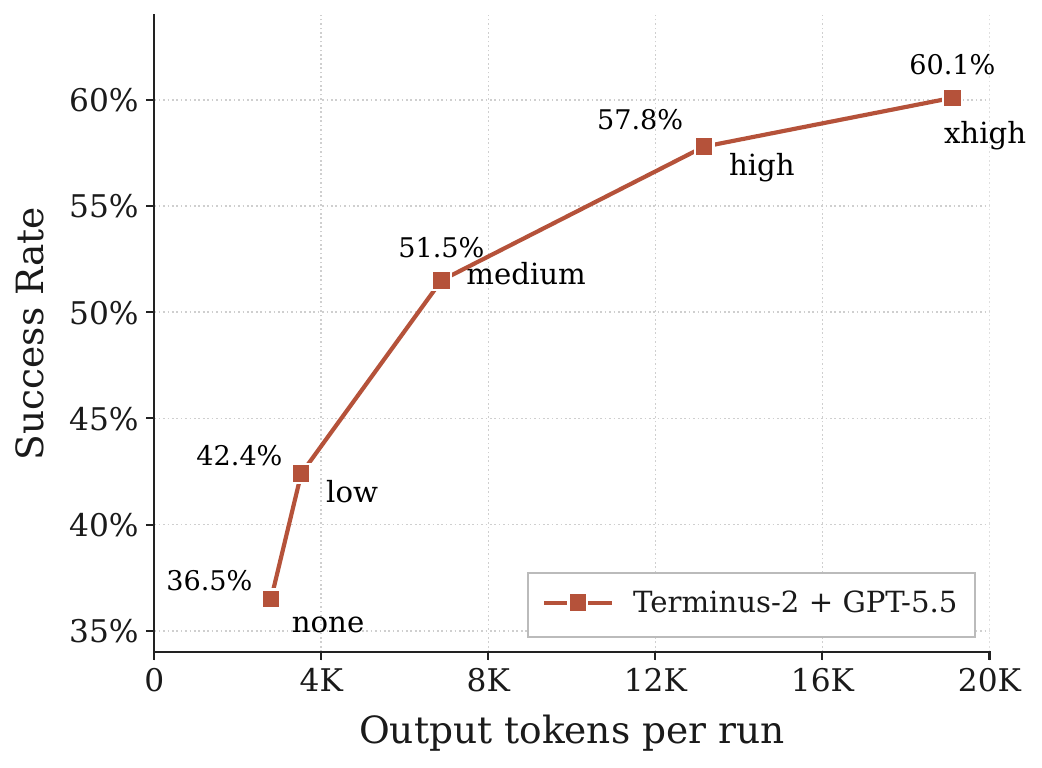}
  \caption{\textbf{Thinking-effort scaling results}. Success rate across five thinking-effort settings for GPT-5.5; higher effort costs more tokens and steadily lifts performance.}
  \label{fig:effort-ablation}
\end{minipage}
\end{figure}

\noindent\textbf{Task-execution time budget.}
\Cref{fig:time-budget-scaling} isolates the effect of the per-task time limit for Terminus-2 + GPT-5.5~(\texttt{xhigh}). As the time limit increases from 150s to 2400s, the number of timed-out trials drops sharply from 337 to 4 out of 600, while the success rate rises from 33.0\% to 60.1\%. This 27.1-point improvement is obtained without changing the model or agent scaffold, indicating that task-execution time is a major factor in measured performance. The close correspondence between fewer timeouts and higher success suggests that many low-budget failures are premature terminations rather than fundamental reasoning failures: the agent is often on a viable trajectory but does not have enough time to complete the task. The gains diminish beyond 1200s, where timeouts are already relatively rare (34 trials) and success reaches 57.1\%. At this point, the remaining failures are more likely to reflect genuine task difficulty rather than truncation. We therefore use a \textbf{2400s} per-task time limit as the default for all other experiments unless otherwise specified, as it nearly eliminates timeouts and provides the cleanest estimate of the agent's task-solving ceiling under this configuration.

\noindent\textbf{Thinking-effort scaling.} \Cref{fig:effort-ablation} shows how Terminus-2 $+$ GPT-5.5 responds to increased reasoning effort. Success rate improves monotonically with the reasoning budget, rising from 36.5\% under the \texttt{none} setting to 60.1\% under \texttt{xhigh}. The largest gains occur at lower effort levels: moving from \texttt{none} to \texttt{medium} recovers roughly 15 percentage points with a modest increase in token cost. Beyond this point, the returns diminish. In particular, increasing effort from \texttt{high} to \texttt{xhigh} improves success by only 2.3 points, while nearly doubling the average number of output tokens from $\sim$13K to $\sim$19K per run. These results indicate that reasoning effort is a reliable lever for improving accuracy, but its cost-effectiveness declines sharply at high effort levels. Medium-to-high settings therefore provide the best accuracy--cost trade-off for most deployments.

\begin{figure}
    \centering
    \includegraphics[width=0.98\linewidth]{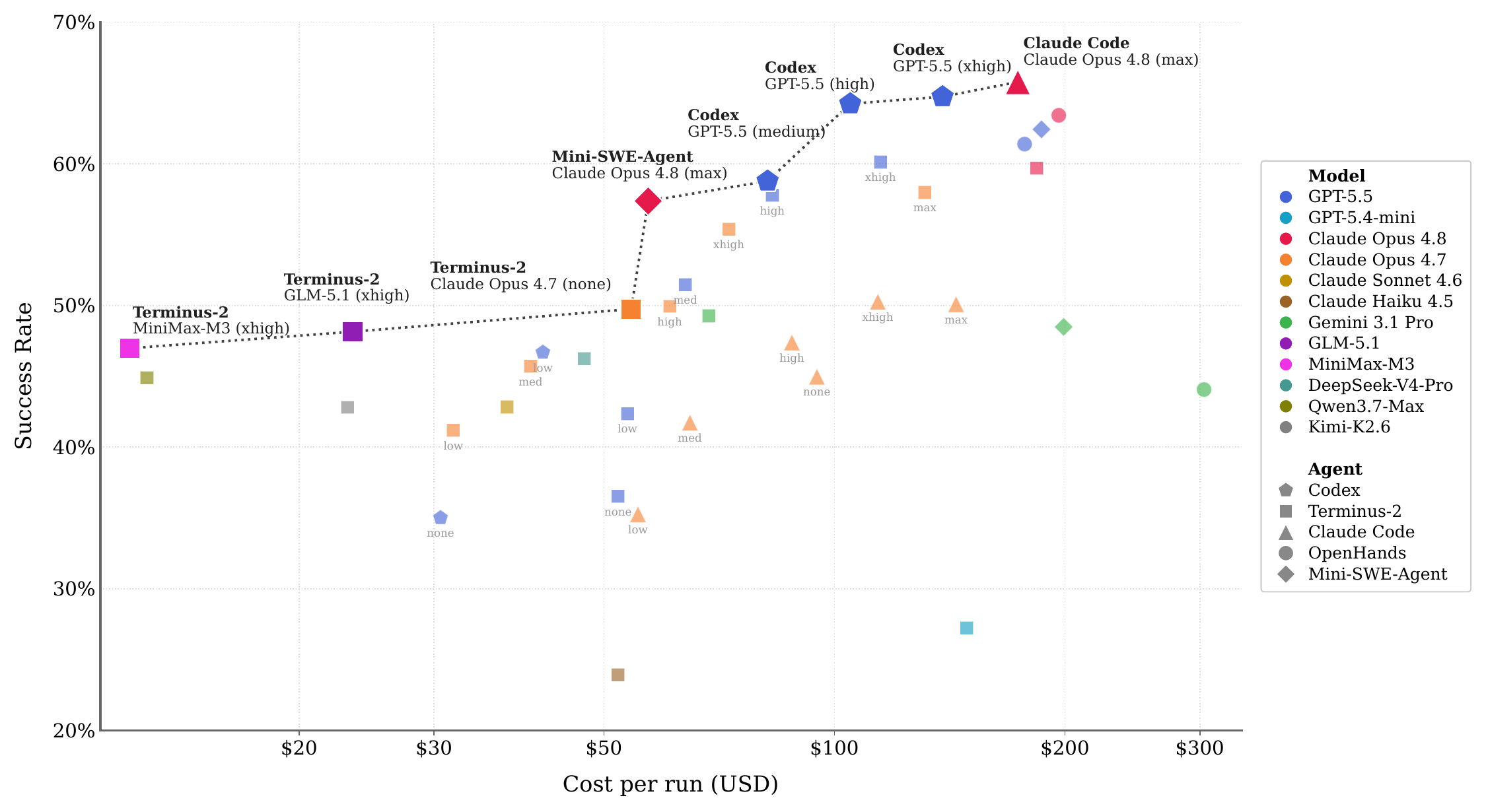}
    \caption{\textbf{Success rate \vs cost per run across agent--model--effort configurations.} Each point is one configuration; the connected line marks the Pareto frontier (best success rate at a given cost). Low-cost efficiency comes from Terminus-2 with open-weight models ($\sim$47--48\% at \$12--23/run), while Claude Code with Claude Opus 4.8 (max) reaches the highest success rate at 65.8\% (\$173.61/run); Codex with GPT-5.5 (\texttt{xhigh}) is close behind at 64.7\% for $\sim$\$138/run. Returns flatten beyond $\sim$\$105/run.}
    \label{fig:success_vs_cost}
    \vspace{-10pt}
\end{figure}
\noindent\textbf{Cost--performance trade-off.} 
\Cref{fig:success_vs_cost} compares task success rate with the average dollar cost per run across 39 agent--model configurations, covering five agent scaffolds and multiple reasoning-effort settings. Overall, cost spans more than an order of magnitude, from roughly \$12 to \$304 per run, while success rates range from 23.9\% to 65.8\%. The highlighted Pareto frontier shows that a small subset of configurations achieves the best trade-off between success and cost. At the low-cost end, Terminus-2 with open-weight models is particularly efficient, with MiniMax-M3 reaching $\sim$47\% success at \$12/run and GLM-5.1 reaching $\sim$48\% at \$23/run. In the mid-cost regime, Terminus-2 with Claude Opus 4.7 and Mini-SWE-Agent with Claude Opus 4.8 further improve the frontier, achieving around 50--57\% success at roughly \$54--\$57 per run. The highest success rates are obtained by Claude Code with Claude Opus 4.8 (\texttt{max}), which reaches 65.8\% at \$173.61/run under maximum reasoning effort.

These results reveal two main trends. First, returns diminish sharply near the top of the frontier: increasing Codex/GPT-5.5 from \texttt{high} to \texttt{xhigh} effort raises success by less than one percentage point while adding about \$33 per run. Second, the scaffold has a substantial impact on cost--performance efficiency. The same model can occupy very different regions of the plot depending on the scaffold; for example, Claude Opus 4.7 achieves comparable accuracy at much lower cost under Terminus-2 than under Claude Code.

\begin{wraptable}{r}{0.46\textwidth}
  \centering
  \small
  \vspace{-\baselineskip}
  \caption{\textbf{Success rate across three open-source agents}. Best per row in \textbf{bold}; $\Delta$ is GPT-5.5~\texttt{xhigh} minus Opus 4.8~\texttt{max}. The stronger model depends on the agent scaffold.}
  \label{tab:model-comparison}
  \begin{tabular}{l c c c}
    \toprule
    Agent & Opus 4.8 & GPT-5.5 & $\Delta$ \\
    \midrule
    Mini-SWE-Agent & $57.4$          & $\mathbf{62.4}$ & $+5.0$ \\
    OpenHands  & $\mathbf{63.4}$ & $61.4$          & $-2.0$ \\
    Terminus-2     & $59.7$          & $\mathbf{60.1}$ & $+0.4$ \\
    \midrule
    Mean           & $60.2$          & $\mathbf{61.3}$ & $+1.1$ \\
    \bottomrule
  \end{tabular}
\end{wraptable}

\noindent\textbf{Agent-dependent model performance.}
\Cref{tab:model-comparison} compares Claude Opus 4.8 and GPT-5.5 across three open-source agent scaffolds, all models are with highest reasoning effort. When averaged over the three harnesses, the two models perform similarly: GPT-5.5 achieves a mean success rate of 61.3\%, only 1.1 percentage points higher than Opus~4.8 at 60.2\%. However, this small aggregate difference masks substantial scaffold-dependent variation. GPT-5.5 outperforms Claude Opus~4.8 by 5.0 points with Mini-SWE-Agent (62.4\% vs.\ 57.4\%), whereas Opus~4.8 leads by 2.0 points with OpenHands (63.4\% vs.\ 61.4\%). Under Terminus-2, the two models are effectively tied, with GPT-5.5 ahead by only 0.4 points (60.1\% vs.\ 59.7\%). These reversals indicate that relative model performance is not invariant to the agent scaffold. Consequently, the choice of harness can have a comparable effect to the choice of underlying model, and conclusions drawn from a single scaffold may misrepresent the relative capabilities of the models.

\begin{figure}[t]
\centering
\includegraphics[width=0.96\linewidth]{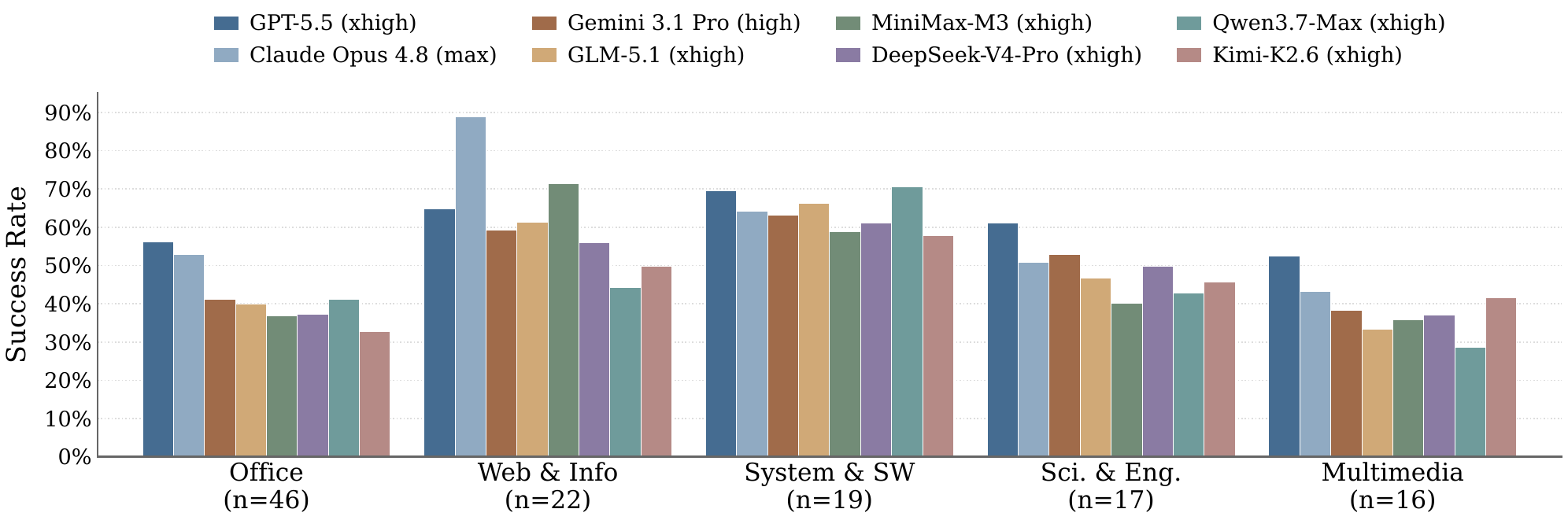}
\vspace{-10pt}
\caption{\textbf{Per-category success rates on TUA-Bench for eight models}, each run with the Terminus-2 agent under the indicated reasoning-effort setting (in parentheses). Bars are grouped by task category, with the number of tasks per category shown below each group (n). A more detailed, task-level breakdown of success rates is provided in \Cref{fig:task-success-heatmap}.}\label{fig:category-success-rate}
\end{figure}

\noindent\textbf{Per-category performance.} \Cref{fig:category-success-rate} reports per-category success rates for eight selected Terminus-2 model configurations (excluding the smallest/legacy variants for readability). Three findings emerge. First, no model dominates uniformly across the suite. GPT-5.5~(\texttt{xhigh}) is the most consistent performer, ranking at or near the top in every category, whereas Claude Opus~4.8~(\texttt{max}) leads by a wide margin on Web \& Info but falls closer to the middle of the field elsewhere, suggesting a more concentrated advantage. Second, task category strongly shapes performance. System \& SW is comparatively tractable, with all models clustered in a narrow high-performing band, while Office and Multimedia are consistently difficult: most models fall below 45\%, and even the strongest systems remain only in the mid-50\% range. Third, absolute success rates remain well below saturation, indicating that TUA-Bench continues to distinguish among frontier systems while leaving substantial headroom. 

\begin{figure}[t]
\centering
\includegraphics[width=0.96\linewidth]{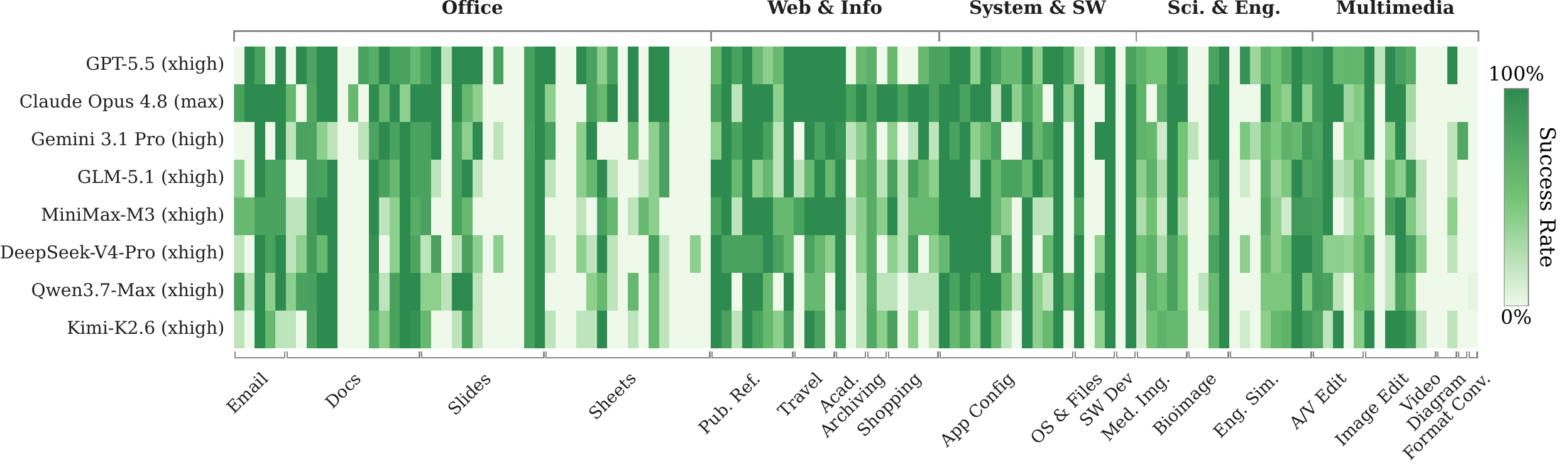}
\vspace{-10pt}
\caption{\textbf{Task-level success rate heatmap.} Mean success rate runs for eight models evaluated with the Terminus-2 agent on each \benchname task. Rows correspond to model configurations and columns correspond to individual tasks, grouped by category and subcategory. Darker cells indicate higher success rate. The heatmap reveals substantial within-category heterogeneity: each category contains both broadly solved tasks and tasks that remain difficult for nearly all models, highlighting task-level capability gaps that are obscured by category-level averages.}\label{fig:task-success-heatmap}
\vspace{-10pt}
\end{figure}

Category-level averages can obscure substantial task-level variation. To expose this structure, \Cref{fig:task-success-heatmap} shows mean success rates for individual tasks, grouped by category and subcategory. The heatmap reveals strong within-category heterogeneity: many categories contain both broadly solved tasks and tasks that nearly all models fail, so moderate averages often reflect a mixture of easy and hard tasks rather than uniform partial success. This pattern is especially visible in Multimedia and parts of Office, where broad low-reward regions indicate shared capability gaps. The heatmap also supports the aggregate trends, with GPT-5.5 and Claude Opus~4.8 appearing darker overall, and shows that Opus~4.8's Web \& Info advantage is concentrated in specific task groups. Together, \Cref{fig:category-success-rate,fig:task-success-heatmap} suggest that progress on TUA-Bench requires addressing the specific hard tasks that current agents consistently fail, not only improving aggregate category-level performance.

\section{Limitations}\label{sec:limitations-detail}

\benchname focuses on terminal-based computer use and therefore does not cover the full spectrum of computer interaction. Some applications still lack mature CLI or headless support, limiting the workflows that can be faithfully represented, though the growing availability of CLI tools may make this setting increasingly relevant. The professional track covers only a limited sample of specialized domains, and all task descriptions are currently English-only. Finally, public release may expose tasks to future model training data, requiring periodic benchmark refreshes, while fixed headless tool versions require ongoing container maintenance.

\section{Conclusion}\label{sec:conclusion}

We introduced \benchname, a benchmark for evaluating general-purpose terminal-use agents across diverse everyday and professional workflows. Complementing prior computer-use benchmarks that focus primarily on GUI interaction or programming-centric terminal tasks, \benchname targets text-native command-line environments, where agents must plan, invoke tools, monitor execution, and recover from errors through terminal interaction alone. The benchmark contains 120 manually curated tasks spanning routine digital work and domain-specific workflows co-designed with experts across multiple fields. Our evaluation of frontier models and agent frameworks shows that even the strongest configuration achieves only a 65.8\% success rate, highlighting reliable terminal-based computer use as a challenging open problem. We open-source \benchname to enable reproducible evaluation, lower the barrier for developing new terminal-use agents, and support community-driven progress toward reliable computer-use systems.

\clearpage
\bibliographystyle{assets/plainnat}
\bibliography{paper}

\clearpage
\beginappendix
\section{Additional Results}

\subsection{Full Results and Online Leaderboard}
The complete results are reported in \Cref{tab:full-results}, where all configurations are ranked by success rate. In addition, we maintain a public leaderboard at \url{https://tuabench.ai}, which will be continuously updated with evaluations of newly released models.

\subsection{Task-Level Results}
\begin{figure}[t]
    \centering
    \includegraphics[width=0.55\linewidth]{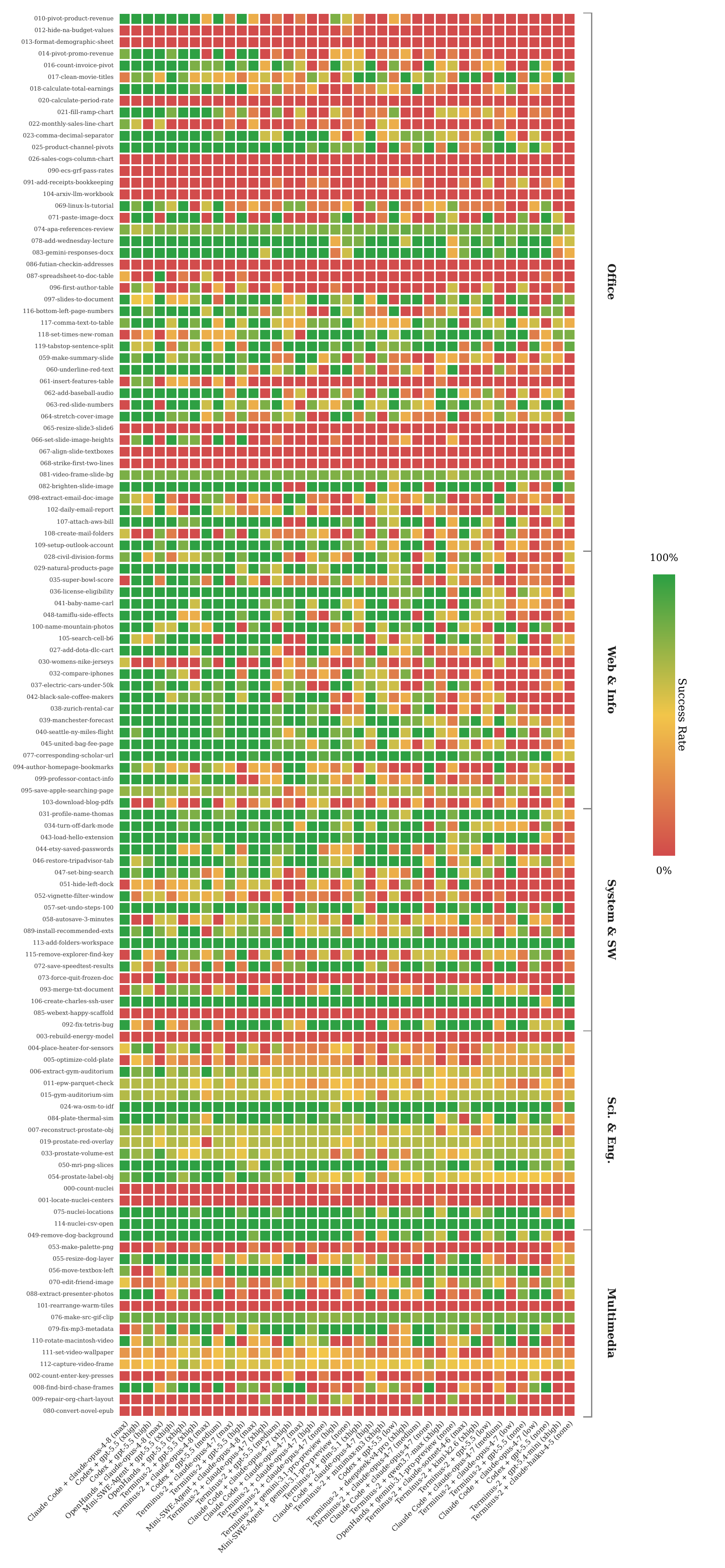}
    \caption{\textbf{Full task-level results on TUA-Bench across agent–model combinations.} Each row is an individual task (identifier and short name on the left), grouped by the five top-level categories. Each column is an agent paired with a specific model under a given reasoning-effort setting (bottom labels, in the form agent + model (effort)). Cell color encodes mean reward across runs, from red (0.0\%) to green (100.0\%). Best viewed in color and zoomed in.}
    \label{fig:app-task-resolution-heatmap}
\end{figure}

\noindent\textbf{Persistent failure modes and the role of the agent scaffold.}
\Cref{fig:app-task-resolution-heatmap} shows the complete per-task reward matrix across all agent--model--effort configurations, providing the most fine-grained view of benchmark performance. Two patterns are particularly salient. First, several tasks appear as near-horizontal red bands that persist across almost all configurations, including multiple chart- and slide-layout tasks in Office (e.g., 065-resize-slide3-slide6, 066-set-slide-image-heights, 067-align-slide-textboxes, and 068-strike-first-two-lines) as well as parts of Multimedia. These patterns indicate failures that are shared across models and scaffolds, suggesting task-intrinsic difficulty rather than configuration-specific weakness. Such persistent failures largely explain the low category-level averages reported in the main text and identify the clearest targets for future progress. Second, the matrix reveals substantial variation across configurations, even when the underlying model is held fixed. This confirms that the agent scaffold and its execution settings materially affect performance: a strong model paired with a less effective scaffold can underperform a weaker model under better orchestration. The remaining tasks fall between these extremes, being solved by some configurations but not others, which is precisely the regime in which the benchmark provides the strongest discrimination.

\begin{figure}[t]
    \centering
    \includegraphics[width=0.5\linewidth]{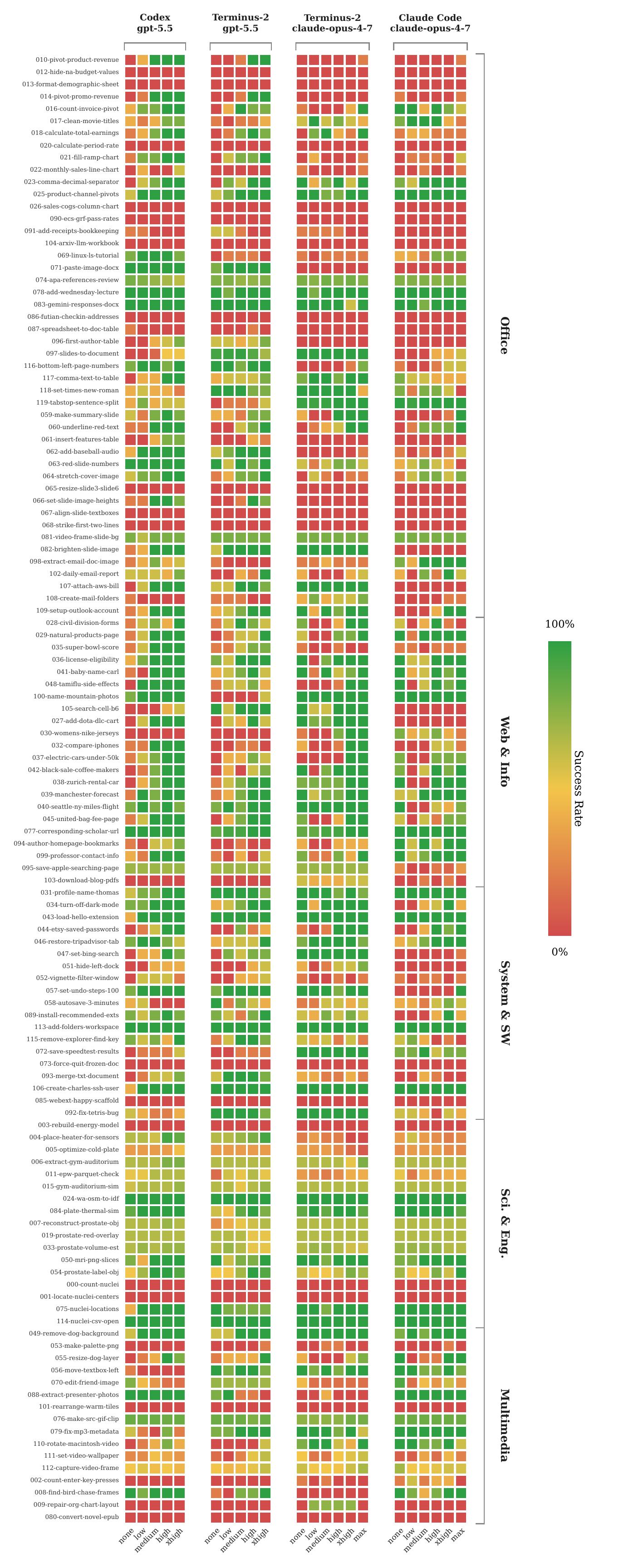}
    \caption{\textbf{Effect of reasoning effort on task-level performance in TUA-Bench.} Rows are individual tasks (identifier and name on the left), grouped by the five task categories (right labels). Columns are organized into four agent–model blocks; within each block, columns correspond to increasing reasoning-effort settings (bottom labels: \texttt{none}, \texttt{low}, \texttt{medium}, \texttt{high}, \texttt{xhigh}, and additionally \texttt{max} for Claude Opus 4.7). Best viewed in color and zoomed in.}
    \label{fig:app-thinking-resolution-heatmap}
\end{figure}

\noindent\textbf{Reasoning effort.}
While \Cref{fig:app-task-resolution-heatmap} jointly varies agent, model, and reasoning effort along the configuration axis, \Cref{fig:app-thinking-resolution-heatmap} isolates the effect of reasoning effort by fixing the agent--model pair within each block and sweeping effort from \texttt{none} to \texttt{xhigh} (and \texttt{max}, where available). Within most blocks, increasing effort generally shifts rewards upward, consistent with the benefit of additional inference-time computation. However, the effect is not uniformly monotonic. Many tasks saturate at medium or high effort, with limited gains from further computation, while others exhibit non-monotonic behavior in which the highest-effort setting performs no better than, or occasionally worse than, an intermediate setting. This suggests diminishing or even negative returns when the limiting factor is not deliberation but missing capability, inadequate tooling, or ineffective execution. Notably, the universally difficult tasks identified in \Cref{fig:app-task-resolution-heatmap} remain unresolved across the full effort sweep, reinforcing that these failures cannot be addressed by reasoning budget alone. Finally, comparisons across blocks at the same effort level show that differences induced by the agent--model pairing are at least as large as those induced by effort scaling, underscoring the need to consider scaffold design and inference-time allocation jointly rather than in isolation.

\begin{table}[t]
\centering
\small
\caption{\textbf{Full results on the \benchname,} ranked by success rate. We evaluate five agent
scaffolds (Claude Code, Codex, OpenHands, Mini-SWE-Agent, Terminus-2) paired
with a range of frontier and open-weight models under varying reasoning-effort
budgets. Success rate is the per-task pass rate
averaged over five independent runs. Pass@$k$ is the fraction of tasks solved within $k$ attempts, and
All-5 is the fraction solved on every attempt. Rows are sorted by success rate;
the best value in each column is shown in \textbf{bold.}}\label{tab:full-results}
\setlength{\tabcolsep}{8pt}
\renewcommand{\arraystretch}{1.05}
\begin{tabular}{l l l c c c c}
\toprule
Agent & Model & Thinking & Success Rate (\%) & Pass@1 & Pass@5 & All-5 \\
\midrule
Claude Code    & Claude Opus 4.8           & \texttt{max}    & \textbf{65.8 $\pm$ 0.7} & \textbf{58.8\%} & 64.2\% & \textbf{51.7\%} \\
Codex          & GPT-5.5                   & \texttt{xhigh}  & 64.7 $\pm$ 0.7 & 57.7\% & \textbf{68.3\%} & 42.5\% \\
Codex          & GPT-5.5                   & \texttt{high}   & 64.2 $\pm$ 0.7 & 57.2\% & 66.7\% & 46.7\% \\
OpenHands      & Claude Opus 4.8           & \texttt{max}    & 63.4 $\pm$ 0.6 & 57.3\% & 67.5\% & 45.0\% \\
Mini-SWE-Agent & GPT-5.5                   & \texttt{xhigh}  & 62.4 $\pm$ 0.8 & 54.2\% & 67.5\% & 40.0\% \\
OpenHands      & GPT-5.5                   & \texttt{xhigh}  & 61.4 $\pm$ 1.0 & 54.0\% & 65.0\% & 38.3\% \\
Terminus-2     & GPT-5.5                   & \texttt{xhigh}  & 60.1 $\pm$ 0.6 & 52.3\% & 64.2\% & 31.7\% \\
Terminus-2     & Claude Opus 4.8           & \texttt{max}    & 59.7 $\pm$ 1.0 & 53.8\% & 62.5\% & 42.5\% \\
Codex          & GPT-5.5                   & \texttt{medium} & 58.8 $\pm$ 1.0 & 51.2\% & 62.5\% & 34.2\% \\
Terminus-2     & Claude Opus 4.7           & \texttt{max}    & 58.0 $\pm$ 0.8 & 51.0\% & 64.2\% & 39.2\% \\
Terminus-2     & GPT-5.5                   & \texttt{high}   & 57.8 $\pm$ 1.7 & 49.8\% & 63.3\% & 32.5\% \\
Mini-SWE-Agent & Claude Opus 4.8           & \texttt{max}    & 57.4 $\pm$ 0.6 & 50.2\% & 64.2\% & 34.2\% \\
Terminus-2     & Claude Opus 4.7           & \texttt{xhigh}  & 55.4 $\pm$ 0.8 & 48.0\% & 59.2\% & 37.5\% \\
Terminus-2     & GPT-5.5                   & \texttt{medium} & 51.5 $\pm$ 1.3 & 42.8\% & 62.5\% & 21.7\% \\
Claude Code    & Claude Opus 4.7           & \texttt{xhigh}  & 50.3 $\pm$ 1.0 & 43.0\% & 57.5\% & 29.2\% \\
Claude Code    & Claude Opus 4.7           & \texttt{max}    & 50.1 $\pm$ 0.6 & 42.8\% & 55.8\% & 30.8\% \\
Terminus-2     & Claude Opus 4.7           & \texttt{high}   & 49.9 $\pm$ 1.2 & 42.5\% & 58.3\% & 25.0\% \\
Terminus-2     & Claude Opus 4.7           & \texttt{none}   & 49.7 $\pm$ 1.1 & 41.7\% & 58.3\% & 27.5\% \\
Terminus-2     & Gemini 3.1 Pro            & \texttt{high}   & 49.3 $\pm$ 1.8 & 41.2\% & 57.5\% & 24.2\% \\
Mini-SWE-Agent & Gemini 3.1 Pro            & \texttt{none}   & 48.5 $\pm$ 1.5 & 40.0\% & 57.5\% & 20.8\% \\
Terminus-2     & GLM-5.1                   & \texttt{xhigh}  & 48.1 $\pm$ 1.3 & 40.3\% & 59.2\% & 20.8\% \\
Claude Code    & Claude Opus 4.7           & \texttt{high}   & 47.4 $\pm$ 0.4 & 40.7\% & 54.2\% & 27.5\% \\
Terminus-2     & MiniMax-M3                & \texttt{xhigh}  & 47.0 $\pm$ 1.3 & 41.2\% & 59.2\% & 22.5\% \\
Codex          & GPT-5.5                   & \texttt{low}    & 46.7 $\pm$ 1.1 & 39.0\% & 61.7\% & 20.0\% \\
Terminus-2     & DeepSeek-V4 Pro           & \texttt{xhigh}  & 46.2 $\pm$ 0.8 & 38.0\% & 57.5\% & 18.3\% \\
Terminus-2     & Claude Opus 4.7           & \texttt{medium} & 45.7 $\pm$ 0.7 & 37.8\% & 51.7\% & 23.3\% \\
Claude Code    & Claude Opus 4.7           & \texttt{none}   & 45.0 $\pm$ 1.1 & 37.7\% & 50.8\% & 24.2\% \\
Terminus-2     & Qwen3.7-Max               & \texttt{xhigh}  & 44.9 $\pm$ 0.7 & 37.7\% & 57.5\% & 21.7\% \\
OpenHands      & Gemini 3.1 Pro            & \texttt{none}   & 44.1 $\pm$ 1.5 & 35.8\% & 56.7\% & 19.2\% \\
Terminus-2     & Claude Sonnet 4.6         & \texttt{max}    & 42.8 $\pm$ 0.3 & 34.8\% & 49.2\% & 20.0\% \\
Terminus-2     & Kimi K2.6                 & \texttt{xhigh}  & 42.8 $\pm$ 1.8 & 35.3\% & 55.8\% & 18.3\% \\
Terminus-2     & GPT-5.5                   & \texttt{low}    & 42.4 $\pm$ 1.7 & 33.5\% & 51.7\% & 13.3\% \\
Claude Code    & Claude Opus 4.7           & \texttt{medium} & 41.7 $\pm$ 0.6 & 34.8\% & 50.8\% & 19.2\% \\
Terminus-2     & Claude Opus 4.7           & \texttt{low}    & 41.2 $\pm$ 1.0 & 32.7\% & 45.8\% & 21.7\% \\
Terminus-2     & GPT-5.5                   & \texttt{none}   & 36.5 $\pm$ 0.8 & 28.2\% & 45.0\% & 14.2\% \\
Claude Code    & Claude Opus 4.7           & \texttt{low}    & 35.3 $\pm$ 0.3 & 28.3\% & 40.8\% & 16.7\% \\
Codex          & GPT-5.5                   & \texttt{none}   & 35.0 $\pm$ 0.9 & 26.2\% & 49.2\% & 8.3\%  \\
Terminus-2     & GPT-5.4 mini              & \texttt{xhigh}  & 27.2 $\pm$ 1.4 & 20.0\% & 41.7\% & 6.7\%  \\
Terminus-2     & Claude Haiku 4.5          & \texttt{none}   & 23.9 $\pm$ 1.5 & 15.7\% & 30.8\% & 3.3\%  \\
\bottomrule
\end{tabular}
\end{table}

\clearpage
\newpage
\section{Full Task List}

The following section enumerates all tasks in the benchmark, specifying the category of each task alongside the corresponding instruction prompt used to guide the terminal-use agent.

\begin{instructionbox}[Scientific \& Engineering - Bioimage Analysis - 000-count-nuclei]
Use the provided microscopy images to count how many nuclei are present.

CellProfiler is available in the environment. Inspect the images and use an appropriate software workflow to estimate the same nuclei count from the nuclear stain image.

Inputs:
- \textasciigrave{}/app/input/images/1-162hrh2ax2.tif\textasciigrave{}
- \textasciigrave{}/app/input/images/1-162hrhoe2.tif\textasciigrave{}

Required final output:
- \textasciigrave{}/app/artifacts/nuclei\_count.txt\textasciigrave{}

Write only the integer nuclei/cell count in the output file.
\end{instructionbox}

\begin{instructionbox}[Scientific \& Engineering - Bioimage Analysis - 001-locate-nuclei-centers]
Use the provided microscopy images to locate the nuclei/cells.

CellProfiler is available in the environment. Inspect the images and use an appropriate software workflow to estimate the center location of each nucleus/cell from the nuclear stain image.

Inputs:
- \textasciigrave{}/app/input/images/1-162hrh2ax2.tif\textasciigrave{}
- \textasciigrave{}/app/input/images/1-162hrhoe2.tif\textasciigrave{}

Required final output:
- \textasciigrave{}/app/artifacts/nuclei\_locations.csv\textasciigrave{}

Write a CSV file with exactly these columns:

\textasciigrave{}\textasciigrave{}\textasciigrave{}csv
x,y
\textasciigrave{}\textasciigrave{}\textasciigrave{}

Each data row should contain one nucleus/cell center in pixel coordinates. The output must be generated from the provided input images.
\end{instructionbox}

\begin{instructionbox}[Multimedia \& Design - Video Understanding - 002-count-enter-key-presses]
The input video is available at \textasciigrave{}/app/input/85229750.mp4\textasciigrave{}.

Inspect the video and count how many times the person hits the Enter key.

Write your answer to \textasciigrave{}/app/result.txt\textasciigrave{} as a single integer, with no extra text.

\textasciigrave{}\textasciigrave{}\textasciigrave{}text
\textless{}count\textgreater{}
\textasciigrave{}\textasciigrave{}\textasciigrave{}
\end{instructionbox}

\begin{instructionbox}[Scientific \& Engineering - Engineering Simulation - 003-rebuild-energy-model]
Given \textasciigrave{}/app/input/floorplan.png\textasciigrave{}, \textasciigrave{}/app/input/weather.epw\textasciigrave{}, and \textasciigrave{}/app/input/task\_plan.json\textasciigrave{}, reconstruct a 3D OpenStudio building model from the floor plan, run an annual EnergyPlus simulation, and write the results under \textasciigrave{}/app/artifacts\textasciigrave{}.

Requirements:

1. Save the reconstructed building model to \textasciigrave{}/app/artifacts/reconstructed\_building.osm\textasciigrave{}.
2. Save the translated EnergyPlus model to \textasciigrave{}/app/artifacts/generated\_building.idf\textasciigrave{}.
3. Write the simulation outputs to \textasciigrave{}/app/artifacts/energyplus\_run\textasciigrave{}.
4. Save review renders under \textasciigrave{}/app/artifacts/render\_views\textasciigrave{} and also write \textasciigrave{}/app/artifacts/building\_render.png\textasciigrave{}.
5. Write \textasciigrave{}/app/artifacts/simulation\_summary.txt\textasciigrave{} with exactly these keys in exactly this order:
   - \textasciigrave{}bldg\_id=\textless{}int\textgreater{}\textasciigrave{}
   - \textasciigrave{}translated\_version=\textless{}string\textgreater{}\textasciigrave{}
   - \textasciigrave{}building\_name=\textless{}string\textgreater{}\textasciigrave{}
   - \textasciigrave{}weather\_file=\textless{}string\textgreater{}\textasciigrave{}
   - \textasciigrave{}row\_count=\textless{}int\textgreater{}\textasciigrave{}
   - \textasciigrave{}annual\_electricity\_kwh=\textless{}float\textgreater{}\textasciigrave{}
   - \textasciigrave{}annual\_natural\_gas\_kwh=\textless{}float\textgreater{}\textasciigrave{}
   - \textasciigrave{}annual\_fuel\_oil\_kwh=\textless{}float\textgreater{}\textasciigrave{}
   - \textasciigrave{}annual\_site\_energy\_kwh=\textless{}float\textgreater{}\textasciigrave{}

Use the \textasciigrave{}bldg\_id\textasciigrave{}, \textasciigrave{}weather\_file\textasciigrave{}, \textasciigrave{}row\_count\textasciigrave{}, meter names, and meter output frequency from \textasciigrave{}/app/input/task\_plan.json\textasciigrave{}.

Constraints:

- Keep all files in \textasciigrave{}/app/input/*\textasciigrave{} unchanged.
- Store all outputs under \textasciigrave{}/app/artifacts\textasciigrave{}.
- The verifier checks annual energy totals against hidden ground truth with \textasciigrave{}1\%\textasciigrave{} relative tolerance.
\end{instructionbox}

\begin{instructionbox}[Scientific \& Engineering - Engineering Simulation - 004-place-heater-for-sensors]
Given \textasciigrave{}/app/input/heater\_design\_request.json\textasciigrave{}, find where to place the fixed-temperature heater so the four sensors match their target temperatures.

Use the installed OpenFOAM tooling or any defensible numerical search workflow. Keep \textasciigrave{}/app/input/*\textasciigrave{} unchanged.

Problem setup:
- Plate dimensions: \textasciigrave{}0.08 m x 0.05 m x 0.005 m\textasciigrave{}
- Heater patch: \textasciigrave{}0.01 m x 0.01 m\textasciigrave{}, fixed at \textasciigrave{}400 K\textasciigrave{}, on the top face
- Bottom and side faces are insulated
- Non-heater top surface weakly cools to \textasciigrave{}300 K\textasciigrave{} with \textasciigrave{}valueFraction = 0.005\textasciigrave{}
- Thermal diffusivity: \textasciigrave{}8.4e-5 m\textasciicircum{}2/s\textasciigrave{}
- Final time: \textasciigrave{}120 s\textasciigrave{}
- The four sensor positions and target temperatures are listed in \textasciigrave{}heater\_design\_request.json\textasciigrave{}
- Match each target sensor temperature within \textasciigrave{}0.5 K\textasciigrave{}

Required outputs:
1. Write \textasciigrave{}/app/artifacts/heater\_placement\_result.json\textasciigrave{} as a nested JSON object with this shape:

\textasciigrave{}\textasciigrave{}\textasciigrave{}json
\{
  "heater\_origin\_m": ["\textless{}x\textgreater{}", "\textless{}y\textgreater{}", "\textless{}z\textgreater{}"],
  "heater\_center\_m": ["\textless{}x\textgreater{}", "\textless{}y\textgreater{}", "\textless{}z\textgreater{}"],
  "predicted\_sensor\_temperatures": [
    \{
      "name": "S1",
      "point\_m": ["\textless{}x\textgreater{}", "\textless{}y\textgreater{}", "\textless{}z\textgreater{}"],
      "temperature\_K": "\textless{}number\textgreater{}",
      "target\_temperature\_K": "\textless{}number\textgreater{}",
      "error\_K": "\textless{}predicted minus target\textgreater{}"
    \}
  ],
  "max\_abs\_error\_K": "\textless{}number\textgreater{}",
  "method\_summary": "\textless{}short description of how you searched or simulated\textgreater{}"
\}
\textasciigrave{}\textasciigrave{}\textasciigrave{}

Use numeric JSON values in your final file, not strings. Include all four sensors, \textasciigrave{}S1\textasciigrave{} through \textasciigrave{}S4\textasciigrave{}.

2. Write \textasciigrave{}/app/artifacts/heater\_placement\_top\_view.svg\textasciigrave{}, a top-view SVG showing the plate outline, the selected heater patch, and the four sensor locations. This is for manual inspection in the artifact folder.
\end{instructionbox}

\begin{instructionbox}[Scientific \& Engineering - Engineering Simulation - 005-optimize-cold-plate]
Given \textasciigrave{}/app/input/task\_plan.json\textasciigrave{}, improve the thermal performance of the liquid-cooled aluminum cold plate by changing only the internal coolant-side geometry inside the plate.

Use the installed OpenFOAM and ParaView tooling from this environment. Keep \textasciigrave{}/app/input/*\textasciigrave{} unchanged. No starter OpenFOAM case is provided. Create your own case and automation under \textasciigrave{}/app/artifacts/improved\_case\textasciigrave{}.

Fixed design conditions:
- External plate envelope: \textasciigrave{}80 mm x 80 mm x 10 mm\textasciigrave{}
- Plate material: aluminum
- Coolant: water
- Heat source footprint: \textasciigrave{}20 mm x 20 mm\textasciigrave{}, centered on the bottom face
- Chip power: \textasciigrave{}300 W\textasciigrave{}
- Coolant inlet temperature: \textasciigrave{}300 K\textasciigrave{}
- Coolant inlet velocity: \textasciigrave{}0.05 m/s\textasciigrave{}
- Flow direction: inlet on the \textasciigrave{}x-min\textasciigrave{} face, outlet on the \textasciigrave{}x-max\textasciigrave{} face
- Dry top face temperature used by the verifier reference physics: fixed at \textasciigrave{}300 K\textasciigrave{}
- Simulation style: simple, defensible steady-state CHT smoke test

What must stay fixed:
- The external plate size and overall bounding box
- The chip footprint, chip location, and chip power
- The coolant type, inlet temperature, inlet velocity, inlet face, outlet face, and overall flow direction
- The non-coolant-side boundary conditions described in \textasciigrave{}task\_plan.json\textasciigrave{} and enforced by the verifier

What is free to change:
- The internal fluid-path geometry inside the plate
- The internal solid-fluid interface geometry inside the channel region
- The internal passive solid/fluid distribution inside the plate, as long as it remains physically plausible

Design constraints:
- Stay fully inside the \textasciigrave{}80 x 80 x 10 mm\textasciigrave{} plate envelope
- Maintain a continuous connected fluid path from inlet to outlet
- Use passive geometry only
- Minimum solid thickness: \textasciigrave{}0.5 mm\textasciigrave{}
- Minimum fluid gap / opening: \textasciigrave{}0.5 mm\textasciigrave{}
- Do not create a numerically fragile or obviously blocked flow path

Success criteria:
- Chip average temperature \textasciigrave{}\textless{}= 335 K\textasciigrave{}
- Pressure drop \textasciigrave{}\textless{}= 12 Pa\textasciigrave{}
- Outlet mass flow \textasciigrave{}\textgreater{}= 0.0012 kg/s\textasciigrave{}

Required outputs:
1. Create \textasciigrave{}/app/artifacts/improved\_case\textasciigrave{} and place your OpenFOAM case, geometry-generation workflow, and automation there.
2. Regenerate the simulation outputs by running your artifact case.
3. Write \textasciigrave{}/app/artifacts/metrics.json\textasciigrave{} with these keys:
   - \textasciigrave{}chip\_average\_temperature\_k\textasciigrave{}
   - \textasciigrave{}solid\_max\_temperature\_k\textasciigrave{}
   - \textasciigrave{}pressure\_drop\_pa\textasciigrave{}
   - \textasciigrave{}outlet\_mass\_flow\_kg\_s\textasciigrave{}
   - \textasciigrave{}thermal\_resistance\_k\_per\_w\textasciigrave{}
4. Write \textasciigrave{}/app/artifacts/baseline\_vs\_improved.md\textasciigrave{} with a short baseline-vs-improved comparison.
5. Write these review renders under \textasciigrave{}/app/artifacts/renders\textasciigrave{}:
   - \textasciigrave{}geometry\_view.svg\textasciigrave{}
   - \textasciigrave{}temperature\_view.svg\textasciigrave{}
   - \textasciigrave{}top\_surface\_temperature.svg\textasciigrave{}
\end{instructionbox}

\begin{instructionbox}[Scientific \& Engineering - Engineering Simulation - 006-extract-gym-auditorium]
Given \textasciigrave{}/app/input/building.osm\textasciigrave{}, create a reduced OpenStudio model that keeps only the gym, auditorium, and gym audience portion of the school.

Inputs:
- \textasciigrave{}/app/input/building.osm\textasciigrave{}: the original whole-school ComStock OpenStudio model

Required final output:
- \textasciigrave{}/app/artifacts/gym\_auditorium\_only.osm\textasciigrave{}
\end{instructionbox}

\begin{instructionbox}[Scientific \& Engineering - Medical Imaging - 007-reconstruct-prostate-obj]
Given \textasciigrave{}/app/input/case\_7f3a9c\_mri.nii.gz\textasciigrave{} and \textasciigrave{}/app/input/task\_plan.json\textasciigrave{}, reconstruct a 3D prostate surface model with the installed 3D Slicer and write outputs under \textasciigrave{}/app/artifacts\textasciigrave{}.

Required outputs:
1. Read \textasciigrave{}/app/input/task\_plan.json\textasciigrave{}.
2. Read the input volume data, identify the prostate anatomy, and segment out only the prostate region.
3. Generate \textasciigrave{}/app/artifacts/prostate\_model.obj\textasciigrave{} as a real non-empty 3D prostate mesh derived from that prostate-only segmentation.
4. Generate \textasciigrave{}/app/artifacts/prostate\_render.png\textasciigrave{} as a rendered PNG view of the generated prostate model.

Requirements:
- Use the real 3D Slicer installation in this environment. Headless use is fine.
- Keep \textasciigrave{}/app/input/*\textasciigrave{} completely unchanged.
- Keep all final outputs under \textasciigrave{}/app/artifacts\textasciigrave{}.
- The input is imaging data, so you must inspect it and create the segmentation yourself instead of assuming the whole volume is already the prostate.
- Segment only the prostate. Do not export the whole scan volume, background, or unrelated anatomy.
- The OBJ must be a valid mesh export of the prostate, not placeholder geometry or an unrelated shape.
\end{instructionbox}

\begin{instructionbox}[Multimedia \& Design - Video Understanding - 008-find-bird-chase-frames]
The input video is available at \textasciigrave{}/app/input/18585469.mp4\textasciigrave{}.

Inspect the video and identify the frame range where the person who is chasing the bird appears.

Write your answer to \textasciigrave{}/app/result.txt\textasciigrave{} using exactly this format:

\textasciigrave{}\textasciigrave{}\textasciigrave{}text
start\_frame=\textless{}frame number\textgreater{}
end\_frame=\textless{}frame number\textgreater{}
\textasciigrave{}\textasciigrave{}\textasciigrave{}

Use frame numbers from the original video, with the first frame numbered \textasciigrave{}0\textasciigrave{}.
\end{instructionbox}

\begin{instructionbox}[Multimedia \& Design - Diagram \& Drawing - 009-repair-org-chart-layout]
The draw.io file at \textasciigrave{}/app/input/org\_chart.drawio\textasciigrave{} contains a corporate organizational chart, but the layout has visible quality issues.

Repair the diagram while preserving the same roles, hierarchy, and department colors. Fix the layout so the role boxes do not overlap and the reporting connectors are clean and readable.

Create the repaired editable draw.io file at \textasciigrave{}/app/corporate\_org\_chart.drawio\textasciigrave{} and export the final PNG preview to \textasciigrave{}/app/corporate\_org\_chart.png\textasciigrave{}.

\textasciigrave{}cli-anything-drawio\textasciigrave{} is installed and can be used if needed.
\end{instructionbox}

\begin{instructionbox}[Office \& Productivity - Spreadsheets - 010-pivot-product-revenue]
Given the file \textasciigrave{}/app/BoomerangSales.xlsx\textasciigrave{}, optionally using LibreOffice Calc if needed, could you help me calculate the revenue in a new column based on the Retail Price sheet (taking into account the product price, quantity, and discount)? Afterward, please generate a Pivot Table in a new sheet (Sheet2) that summarizes the revenue of each product, and save the completed spreadsheet in place at \textasciigrave{}/app/BoomerangSales.xlsx\textasciigrave{}.
\end{instructionbox}

\begin{instructionbox}[Scientific \& Engineering - Engineering Simulation - 011-epw-parquet-check]
Given \textasciigrave{}/app/input/building.osm\textasciigrave{}, \textasciigrave{}/app/input/weather.epw\textasciigrave{}, and \textasciigrave{}/app/input/task\_plan.json\textasciigrave{}, use the installed OpenStudio CLI (\textasciigrave{}openstudio\textasciigrave{}) and the bundled EnergyPlus executable to produce a translated IDF, run the simulation, and summarize the annual energy totals.

Here is the workflow I'd like you to follow:

1. Read \textasciigrave{}/app/input/task\_plan.json\textasciigrave{}.
2. Use the real OpenStudio CLI to translate \textasciigrave{}/app/input/building.osm\textasciigrave{} into \textasciigrave{}/app/artifacts/generated\_building.idf\textasciigrave{}.
3. Append the required \textasciigrave{}Output:Meter\textasciigrave{} objects in the exact order listed in \textasciigrave{}/app/input/task\_plan.json\textasciigrave{}, using the reporting frequency listed there.
4. Run the real EnergyPlus simulator with \textasciigrave{}/app/input/weather.epw\textasciigrave{} and write the run directory to \textasciigrave{}/app/artifacts/energyplus\_run\textasciigrave{}.
5. Parse \textasciigrave{}/app/artifacts/energyplus\_run/eplusout.mtr\textasciigrave{} to compute annual totals in kWh for:
   - electricity
   - natural gas
   - fuel oil
   - total site energy, defined as the sum of those three annual totals
6. Write \textasciigrave{}/app/artifacts/simulation\_summary.txt\textasciigrave{} with exactly the following keys in exactly this order:
   - \textasciigrave{}bldg\_id=\textless{}int\textgreater{}\textasciigrave{}
   - \textasciigrave{}translated\_version=\textless{}string\textgreater{}\textasciigrave{}
   - \textasciigrave{}building\_name=\textless{}string\textgreater{}\textasciigrave{}
   - \textasciigrave{}weather\_file=\textless{}string\textgreater{}\textasciigrave{}
   - \textasciigrave{}row\_count=\textless{}int\textgreater{}\textasciigrave{}
   - \textasciigrave{}annual\_electricity\_kwh=\textless{}float\textgreater{}\textasciigrave{}
   - \textasciigrave{}annual\_natural\_gas\_kwh=\textless{}float\textgreater{}\textasciigrave{}
   - \textasciigrave{}annual\_fuel\_oil\_kwh=\textless{}float\textgreater{}\textasciigrave{}
   - \textasciigrave{}annual\_site\_energy\_kwh=\textless{}float\textgreater{}\textasciigrave{}

Use the values recorded in \textasciigrave{}/app/input/task\_plan.json\textasciigrave{} for \textasciigrave{}bldg\_id\textasciigrave{}, \textasciigrave{}weather\_file\textasciigrave{}, and \textasciigrave{}row\_count\textasciigrave{}. The verifier holds the ground-truth annual totals separately, so the inputs exposed under \textasciigrave{}/app/input\textasciigrave{} do not contain the reference answer.

Please keep the following constraints in mind:

- Use the real \textasciigrave{}openstudio\textasciigrave{} CLI and the real EnergyPlus executable. Avoid fake, hand-written, or placeholder simulation outputs.
- Keep all files in \textasciigrave{}/app/input/*\textasciigrave{} unchanged.
- Ensure \textasciigrave{}/app/artifacts/generated\_building.idf\textasciigrave{}, \textasciigrave{}/app/artifacts/simulation\_summary.txt\textasciigrave{}, \textasciigrave{}/app/artifacts/energyplus\_run/eplusout.mtr\textasciigrave{}, \textasciigrave{}/app/artifacts/energyplus\_run/eplusout.sql\textasciigrave{}, and \textasciigrave{}/app/artifacts/energyplus\_run/eplustbl.htm\textasciigrave{} all exist and are non-empty.
- Store the final outputs under \textasciigrave{}/app/artifacts\textasciigrave{} so they are preserved as task artifacts.
\end{instructionbox}

\begin{instructionbox}[Office \& Productivity - Spreadsheets - 012-hide-na-budget-values]
Given the file \textasciigrave{}/app/Date\_Budget\_Variance\_HideNA.xlsx\textasciigrave{}, could you help me manage some missing data? Currently, some missing values are temporarily filled with 'N/A'. Please hide these 'N/A' entries in the table for now, optionally using LibreOffice Calc if needed. Make sure not to delete any cells, and please note that a filter is not needed. Finally, save the completed spreadsheet in place at \textasciigrave{}/app/Date\_Budget\_Variance\_HideNA.xlsx\textasciigrave{}.
\end{instructionbox}

\begin{instructionbox}[Office \& Productivity - Spreadsheets - 013-format-demographic-sheet]
Given the file \textasciigrave{}/app/DemographicProfile.xlsx\textasciigrave{}, could you help me update it, optionally using LibreOffice Calc if needed? Please create a new sheet named "Sheet2". In this new sheet, merge cells A1:C1 and write "Demographic Profile" with a blue (\#0000ff) fill and bold white text. Next, I'd like you to create three pivot tables showing the percentages of Sex, Civil Status, and Highest Educational Attainment. Please stack them one by one in Sheet2, separating each table with a blank line. Once you're done, please save the completed spreadsheet in place at \textasciigrave{}/app/DemographicProfile.xlsx\textasciigrave{}.
\end{instructionbox}

\begin{instructionbox}[Office \& Productivity - Spreadsheets - 014-pivot-promo-revenue]
Given the file \textasciigrave{}/app/EntireSummerSales.xlsx\textasciigrave{}, could you help me summarize the total revenue for each promotion type in a new sheet (Sheet2) with the promotion names as the column headers using the Pivot Table feature? You can optionally use LibreOffice Calc if needed. Once you're done, please save the completed spreadsheet in place at \textasciigrave{}/app/EntireSummerSales.xlsx\textasciigrave{}.
\end{instructionbox}

\begin{instructionbox}[Scientific \& Engineering - Engineering Simulation - 015-gym-auditorium-sim]
Given \textasciigrave{}/app/input/building.osm\textasciigrave{}, run an EnergyPlus simulation for only the gym, auditorium, and gym audience portion of the school.

Inputs:
- \textasciigrave{}/app/input/building.osm\textasciigrave{}: the school ComStock OpenStudio model
- \textasciigrave{}/app/input/weather.epw\textasciigrave{}: the weather file to use for the simulation

Required final outputs:
- \textasciigrave{}/app/artifacts/gym\_auditorium\_only.osm\textasciigrave{}: the reduced OpenStudio model you simulated
- \textasciigrave{}/app/artifacts/generated\_building.idf\textasciigrave{}: the IDF translated from that reduced model
- \textasciigrave{}/app/artifacts/energyplus\_run/eplusout.sql\textasciigrave{}: the EnergyPlus SQL output from the scoped simulation
- \textasciigrave{}/app/artifacts/energyplus\_run/eplusout.end\textasciigrave{}
- \textasciigrave{}/app/artifacts/energyplus\_run/eplusout.err\textasciigrave{}
\end{instructionbox}

\begin{instructionbox}[Office \& Productivity - Spreadsheets - 016-count-invoice-pivot]
Given the file \textasciigrave{}/app/Invoices.xlsx\textasciigrave{}, could you help me create a Pivot Table in a new sheet (Sheet2) to count how many times each "Invoice No." appears, optionally using LibreOffice Calc if needed? Make the pivot count the appearances of the "Invoice No." field directly rather than using another column as a proxy for row count. Please save the completed spreadsheet in place at \textasciigrave{}/app/Invoices.xlsx\textasciigrave{}.
\end{instructionbox}

\begin{instructionbox}[Office \& Productivity - Spreadsheets - 017-clean-movie-titles]
Given the file \textasciigrave{}/app/Movie\_title\_garbage\_clean.xlsx\textasciigrave{}, optionally using LibreOffice Calc if needed, could you help me copy the movie titles from the 'Garbage Movie Titles' column to the 'Clean Movie Titles' column? While doing so, please remove any extra whitespaces and canonicalize the letter cases by capitalizing the first letter of each word and leaving all other letters lowercase. Please make sure not to touch any irrelevant regions, even if they are blank. Once finished, save the completed spreadsheet in place at \textasciigrave{}/app/Movie\_title\_garbage\_clean.xlsx\textasciigrave{}.
\end{instructionbox}

\begin{instructionbox}[Office \& Productivity - Spreadsheets - 018-calculate-total-earnings]
Given the file \textasciigrave{}/app/Multiply\_Time\_Number.xlsx\textasciigrave{}, could you help me figure out my total earnings? You can optionally use LibreOffice Calc if needed. I have calculated the total work hours from the daily hours, and I have an hourly rate. I want to multiply the total hours by the hourly rate to get the total earned amount. However, I can't get the correct answer by directly multiplying the two cells because the "total hours" is formatted as time, while the "hourly rate" is just a number. Could you help me fill in the cell with the correct product? Please don't touch any irrelevant blank regions, and save the completed spreadsheet in place at \textasciigrave{}/app/Multiply\_Time\_Number.xlsx\textasciigrave{}.
\end{instructionbox}

\begin{instructionbox}[Scientific \& Engineering - Medical Imaging - 019-prostate-red-overlay]
Given \textasciigrave{}/app/input/case\_7f3a9c\_mri.nii.gz\textasciigrave{}, create a new NIfTI volume with the prostate overlaid in red.

Required output:
- \textasciigrave{}/app/artifacts/prostate\_red\_overlay.nii.gz\textasciigrave{}

Requirements:
- Use the installed 3D Slicer tooling. Headless use is fine.
- Inspect the MRI data and estimate the prostate region; do not use placeholder geometry.
- The output must be a real \textasciigrave{}.nii.gz\textasciigrave{} NIfTI file with red pixels/voxels marking the prostate region.
- Keep \textasciigrave{}/app/input/*\textasciigrave{} unchanged.
- Keep final outputs under \textasciigrave{}/app/artifacts\textasciigrave{}.
\end{instructionbox}

\begin{instructionbox}[Office \& Productivity - Spreadsheets - 020-calculate-period-rate]
Given the file \textasciigrave{}/app/PeriodRate.xlsx\textasciigrave{}, could you help me calculate the period rate for my data? You can optionally use LibreOffice Calc if needed. Please place the calculations in a new column with the header "Period Rate (\%)", convert the results to a number type, and highlight the highest result with a green (\#00ff00) font. When you are finished, please save the completed spreadsheet in place at \textasciigrave{}/app/PeriodRate.xlsx\textasciigrave{}.
\end{instructionbox}

\begin{instructionbox}[Office \& Productivity - Spreadsheets - 021-fill-ramp-chart]
Given the file \textasciigrave{}/app/RampUpAndDown.xlsx\textasciigrave{}, could you help me with some updates? You can optionally use LibreOffice Calc if needed. I have computed the acceleration in row 2, and I would like you to fill out the remaining rows for columns B and D. Next, please concatenate the values from columns A to D, including their headers (using the pattern "Header: cell value, ..., Header: cell value"), into a new column named "Combined Data" for all rows. In this new column, make sure to keep only 2 decimal digits. Once you are finished, please save the completed spreadsheet in place at \textasciigrave{}/app/RampUpAndDown.xlsx\textasciigrave{}.
\end{instructionbox}

\begin{instructionbox}[Office \& Productivity - Spreadsheets - 022-monthly-sales-line-chart]
Given the file \textasciigrave{}/app/SalesRep.xlsx\textasciigrave{}, optionally using LibreOffice Calc if needed, could you help me work out the monthly total sales in a new row called "Total"? Once calculated, please create a line chart to show the results with the months on the x-axis, and save the completed spreadsheet in place at \textasciigrave{}/app/SalesRep.xlsx\textasciigrave{}.
\end{instructionbox}

\begin{instructionbox}[Office \& Productivity - Spreadsheets - 023-comma-decimal-separator]
Given the file \textasciigrave{}/app/Set\_Decimal\_Separator\_Dot.xlsx\textasciigrave{}, could you help me set the decimal separator as a comma (,) for localized data representation and clarity in visualization? You can optionally use LibreOffice Calc if needed. Please update all the numbers in the sheet while keeping the decimal numbers as-is, and then save the completed spreadsheet in place at \textasciigrave{}/app/Set\_Decimal\_Separator\_Dot.xlsx\textasciigrave{}.
\end{instructionbox}

\begin{instructionbox}[Scientific \& Engineering - Engineering Simulation - 024-wa-osm-to-idf]
Given \textasciigrave{}/app/input/building.osm\textasciigrave{}, \textasciigrave{}/app/input/building\_timeseries.parquet\textasciigrave{}, and \textasciigrave{}/app/input/task\_plan.json\textasciigrave{}, use the installed OpenStudio CLI (\textasciigrave{}openstudio\textasciigrave{}) to produce a translated IDF and a deterministic summary.

Here is the workflow I'd like you to follow:

1. Read \textasciigrave{}/app/input/task\_plan.json\textasciigrave{}.
2. Use the real OpenStudio CLI to translate \textasciigrave{}/app/input/building.osm\textasciigrave{} into \textasciigrave{}/app/generated\_building.idf\textasciigrave{}.
3. Parse the generated IDF to extract:
   - the translated EnergyPlus version from the \textasciigrave{}Version\textasciigrave{} object
   - the building name from the \textasciigrave{}Building\textasciigrave{} object
4. Write \textasciigrave{}/app/translation\_summary.txt\textasciigrave{} with exactly the following keys in exactly this order:
   - \textasciigrave{}bldg\_id=\textless{}int\textgreater{}\textasciigrave{}
   - \textasciigrave{}translated\_version=\textless{}string\textgreater{}\textasciigrave{}
   - \textasciigrave{}building\_name=\textless{}string\textgreater{}\textasciigrave{}
   - \textasciigrave{}row\_count=\textless{}int\textgreater{}\textasciigrave{}
   - \textasciigrave{}annual\_site\_energy\_kwh=\textless{}float\textgreater{}\textasciigrave{}
   - \textasciigrave{}peak\_hourly\_site\_energy\_kwh=\textless{}float\textgreater{}\textasciigrave{}

Use the values recorded in \textasciigrave{}/app/input/task\_plan.json\textasciigrave{} for \textasciigrave{}bldg\_id\textasciigrave{}, \textasciigrave{}row\_count\textasciigrave{}, \textasciigrave{}annual\_site\_energy\_kwh\textasciigrave{}, and \textasciigrave{}peak\_hourly\_site\_energy\_kwh\textasciigrave{}. Those values were precomputed from \textasciigrave{}/app/input/building\_timeseries.parquet\textasciigrave{}, so you do not need to install a parquet reader.

Please keep the following constraints in mind:

- Use the real \textasciigrave{}openstudio\textasciigrave{} CLI. Avoid fake, hand-written, or placeholder IDF output.
- Keep all files in \textasciigrave{}/app/input/*\textasciigrave{} unchanged.
- Ensure both \textasciigrave{}/app/generated\_building.idf\textasciigrave{} and \textasciigrave{}/app/translation\_summary.txt\textasciigrave{} exist and are non-empty.
\end{instructionbox}

\begin{instructionbox}[Office \& Productivity - Spreadsheets - 025-product-channel-pivots]
Given the file \textasciigrave{}/app/SummerSales.xlsx\textasciigrave{}, could you help me create two native pivot tables in a new sheet named "Sheet2" showing the total revenue for each product and sales channel? Use LibreOffice Calc's pivot table/DataPilot feature so the saved workbook contains actual pivot table objects, not manually typed summaries or ordinary formatted tables. Once you're finished, please save the completed spreadsheet in place at \textasciigrave{}/app/SummerSales.xlsx\textasciigrave{}.
\end{instructionbox}

\begin{instructionbox}[Office \& Productivity - Spreadsheets - 026-sales-cogs-column-chart]
Given the file \textasciigrave{}/app/WeeklySales.xlsx\textasciigrave{}, could you help me create a clustered column chart showing the Sales and COGS data for each week in a new sheet named "Sheet2"? You can optionally use LibreOffice Calc if needed. Please set the chart title to "Sales \& COGS", and save the completed spreadsheet in place at \textasciigrave{}/app/WeeklySales.xlsx\textasciigrave{}.
\end{instructionbox}

\begin{instructionbox}[Web \& Information - Shopping \& Commerce - 027-add-dota-dlc-cart]
Given the websites \textasciigrave{}https://www.dota2.com/home\textasciigrave{} and \textasciigrave{}https://store.steampowered.com/\textasciigrave{}, could you help me find the Dota 2 game on Steam and add all of its DLC to the cart? You can do this optionally using Chrome if needed. Open the relevant page and, when finished, leave that exact page selected as the active tab in the foreground window.
\end{instructionbox}

\begin{instructionbox}[Web \& Information - Public Reference - 028-civil-division-forms]
Given the website \textasciigrave{}https://www.justice.gov/\textasciigrave{}, could you help me browse the list of Civil Division forms? You can do this optionally using Chrome if needed. Open the relevant page and, when finished, leave that exact page selected as the active tab in the foreground window.
\end{instructionbox}

\begin{instructionbox}[Web \& Information - Public Reference - 029-natural-products-page]
Given the website \textasciigrave{}https://www.drugs.com/\textasciigrave{}, could you help me browse to the natural products database page? You can do this optionally using Chrome if needed. Open the relevant page and, when finished, leave that exact page selected as the active tab in the foreground window.
\end{instructionbox}

\begin{instructionbox}[Web \& Information - Shopping \& Commerce - 030-womens-nike-jerseys]
Given the website \textasciigrave{}https://www.nba.com/\textasciigrave{}, could you help me browse to a page showing women's Nike jerseys priced over \$60? You can do this optionally using Chrome if needed. Open the relevant page and, when finished, leave that exact page selected as the active tab in the foreground window.
\end{instructionbox}

\begin{instructionbox}[System \& Software Operations - Application \& Environment Config - 031-profile-name-thomas]
Given the Chrome profile settings, lately I have changed my English name to Thomas. I want to update my username. Could you help me change the profile name to \textasciigrave{}Thomas\textasciigrave{}? You can optionally use Chrome if needed, but please make sure the updated profile is successfully written to disk.
\end{instructionbox}

\begin{instructionbox}[Web \& Information - Shopping \& Commerce - 032-compare-iphones]
Given the website \textasciigrave{}https://www.apple.com/\textasciigrave{}, could you help me compare the iPhone 15 Pro Max with the iPhone 14 Pro Max and the iPhone 13 Pro Max? Use Chrome to open Apple's iPhone comparison page for those three models. Open the relevant page and, when finished, leave that exact page selected as the active tab in the foreground window.
\end{instructionbox}

\begin{instructionbox}[Scientific \& Engineering - Medical Imaging - 033-prostate-volume-est]
Given \textasciigrave{}/app/input/case\_7f3a9c\_mri.nii.gz\textasciigrave{}, estimate the volume of the prostate.

Required output:
- \textasciigrave{}/app/artifacts/prostate\_volume.txt\textasciigrave{}

Output format:
- The file must contain just one number.
- The number must be the prostate volume in cubic centimeters, equivalent to milliliters.

Requirements:
- Use the installed 3D Slicer tooling. Headless use is fine.
- Inspect the MRI data and estimate the prostate region; do not use placeholder values.
- Keep \textasciigrave{}/app/input/*\textasciigrave{} unchanged.
- Keep final outputs under \textasciigrave{}/app/artifacts\textasciigrave{}.
\end{instructionbox}

\begin{instructionbox}[System \& Software Operations - Application \& Environment Config - 034-turn-off-dark-mode]
Could you assist me in turning off the dark mode feature? I've noticed that while dark mode is great for reducing glare, it actually makes it more challenging for me to read text clearly, especially with my astigmatism. Dark mode is currently enabled, and you can complete this task optionally using Google Chrome if needed. Please turn off the dark mode from the built-in settings. Open the relevant page and, when finished, leave that exact page selected as the active tab in the foreground window.
\end{instructionbox}

\begin{instructionbox}[Web \& Information - Public Reference - 035-super-bowl-score]
Given the website \textasciigrave{}https://www.nfl.com/\textasciigrave{}, could you help me find the score record for the Super Bowl of the 2019 NFL season (played in 2020) on the NFL website, optionally using Chrome if needed? Open the relevant page and, when finished, leave that exact page selected as the active tab in the foreground window.
\end{instructionbox}

\begin{instructionbox}[Web \& Information - Public Reference - 036-license-eligibility]
Given the website \textasciigrave{}https://www.dmv.virginia.gov/\textasciigrave{}, could you help me find the Driver License Eligibility Requirements? Optionally using Chrome if needed. Open the relevant page and, when finished, leave that exact page selected as the active tab in the foreground window.
\end{instructionbox}

\begin{instructionbox}[Web \& Information - Shopping \& Commerce - 037-electric-cars-under-50k]
Given the website \textasciigrave{}https://www.cars.com/\textasciigrave{}, could you help me find electric cars with a maximum price of \$50,000 within 50 miles of ZIP code 10001? You can do this optionally using Chrome if needed. Open the relevant page and, when finished, leave that exact page selected as the active tab in the foreground window.
\end{instructionbox}

\begin{instructionbox}[Web \& Information - Travel \& Local - 038-zurich-rental-car]
Given the website \textasciigrave{}https://www.rentalcars.com/\textasciigrave{}, could you help me search for a large car from next Monday to Friday with both pick-up and drop-off in Zurich? Please make sure to sort the results by price. You can complete this task optionally using Chrome if needed. Open the relevant page and, when finished, leave that exact page selected as the active tab in the foreground window.
\end{instructionbox}

\begin{instructionbox}[Web \& Information - Travel \& Local - 039-manchester-forecast]
Given the website \textasciigrave{}https://www.accuweather.com/\textasciigrave{}, could you help me find the monthly forecast for Manchester, GB for this month, optionally using Chrome if needed? Open the relevant page and, when finished, leave that exact page selected as the active tab in the foreground window.
\end{instructionbox}

\begin{instructionbox}[Web \& Information - Travel \& Local - 040-seattle-ny-miles-flight]
Given the website \textasciigrave{}https://www.delta.com/\textasciigrave{}, could you help me find flights from Seattle to New York on the 5th of next month and only show those that can be purchased with miles? You may optionally use Chrome if needed. Please use the existing Delta tab. Open the relevant page and, when finished, leave that exact page selected as the active tab in the foreground window.
\end{instructionbox}

\begin{instructionbox}[Web \& Information - Public Reference - 041-baby-name-carl]
Given the website \textasciigrave{}https://www.babycenter.com/child\textasciigrave{}, could you help me find similar names to the name Carl? You can optionally use Chrome if needed. Open the relevant page and, when finished, leave that exact page selected as the active tab in the foreground window.
\end{instructionbox}

\begin{instructionbox}[Web \& Information - Shopping \& Commerce - 042-black-sale-coffee-makers]
Given the website \textasciigrave{}https://shopping.google.com/\textasciigrave{}, could you help me create a list of drip coffee makers that are on sale, priced between \$25 and \$60, and have a black finish? You can achieve this by searching for \textasciigrave{}drip coffee maker\textasciigrave{} and applying the \textasciigrave{}Black\textasciigrave{}, \textasciigrave{}\$25 - \$60\textasciigrave{}, and \textasciigrave{}On sale\textasciigrave{} filters, optionally using Chrome if needed. Open the relevant page and, when finished, leave that exact page selected as the active tab in the foreground window.
\end{instructionbox}

\begin{instructionbox}[System \& Software Operations - Application \& Environment Config - 043-load-hello-extension]
Given the file \textasciigrave{}\textasciitilde{}/Desktop/helloExtension.zip\textasciigrave{}, could you help me unzip it so that the unpacked extension directory is exactly \textasciigrave{}\textasciitilde{}/Desktop/helloExtension\textasciigrave{}? Load \textasciigrave{}\textasciitilde{}/Desktop/helloExtension\textasciigrave{} specifically in Chrome's existing default profile, not a nested extracted subdirectory or a separate temporary profile. The \textasciigrave{}chrome://extensions/\textasciigrave{} page may already be open. When you're done with the setup, please close the browser so the saved profile is written to disk.
\end{instructionbox}

\begin{instructionbox}[System \& Software Operations - Application \& Environment Config - 044-etsy-saved-passwords]
Could you help me navigate to the area in my browser settings where my passwords are stored? I want to check my login information for Etsy without revealing it just yet. You can do this optionally using Chrome if needed. Please navigate to the relevant Chrome settings page and do not reveal any stored passwords. Open the relevant page and, when finished, leave that exact page selected as the active tab in the foreground window.
\end{instructionbox}

\begin{instructionbox}[Web \& Information - Travel \& Local - 045-united-bag-fee-page]
Given the website \textasciigrave{}https://www.united.com/en/us\textasciigrave{}, could you help me navigate to the United Airlines checked bag fee calculator page, optionally using Chrome if needed? Open the relevant page and, when finished, leave that exact page selected as the active tab in the foreground window.
\end{instructionbox}

\begin{instructionbox}[System \& Software Operations - Application \& Environment Config - 046-restore-tripadvisor-tab]
Could you help me bring back the last tab I shut down on my computer? The browser profile is already prepared with \textasciigrave{}https://www.lonelyplanet.com\textasciigrave{} and \textasciigrave{}https://www.airbnb.com\textasciigrave{} currently open, and \textasciigrave{}https://www.tripadvisor.com\textasciigrave{} as the most recently closed tab. Please restore the closed TripAdvisor tab so all three sites are open again, optionally using Chrome if needed. Make sure to leave the browser running when you are done.
\end{instructionbox}

\begin{instructionbox}[System \& Software Operations - Application \& Environment Config - 047-set-bing-search]
Could you help me make Bing the main search engine when I look stuff up on the internet? Please make this change in Chrome's saved Default profile, editing the profile files directly if that is the most reliable way to do it, and then quit the browser so the updated profile is written to disk.
\end{instructionbox}

\begin{instructionbox}[Web \& Information - Public Reference - 048-tamiflu-side-effects]
Given the website \textasciigrave{}https://www.drugs.com/\textasciigrave{}, could you help me show the side effects of Tamiflu? You can optionally use Chrome if needed. Open the relevant page and, when finished, leave that exact page selected as the active tab in the foreground window.
\end{instructionbox}

\begin{instructionbox}[Multimedia \& Design - Image Editing - 049-remove-dog-background]
Given the file \textasciigrave{}/app/dog\_with\_background.png\textasciigrave{}, could you help me make the background transparent, optionally using GIMP if needed? Please keep the original canvas size, preserve the dog subject, fully clear the removed background so no original background color remains in transparent regions, and save the result as \textasciigrave{}/app/dog\_without\_background.png\textasciigrave{}.
\end{instructionbox}

\begin{instructionbox}[Scientific \& Engineering - Medical Imaging - 050-mri-png-slices]
Given \textasciigrave{}/app/input/case\_7f3a9c\_mri.nii.gz\textasciigrave{} and \textasciigrave{}/app/input/task\_plan.json\textasciigrave{}, export every axial T2 layer of the MRI volume as a PNG image.

Required output directory:
- \textasciigrave{}/app/artifacts/png\_slices\textasciigrave{}

Requirements:
- Use the installed 3D Slicer tooling. Headless use is fine.
- The input NIfTI contains two channels/volumes; export the first one as the T2 series.
- Preserve the axial display orientation used by Slicer. If you use array operations directly, this is equivalent to rotating each \textasciigrave{}volume[:, :, z]\textasciigrave{} slice 90 degrees counterclockwise before saving it.
- Use a consistent grayscale window/normalization across the exported slices so the anatomy remains visible.
- Export exactly 15 PNG files, one for each axial slice.
- Use these filenames exactly: \textasciigrave{}prostate\_00\_t2\_slice\_00.png\textasciigrave{}, \textasciigrave{}prostate\_00\_t2\_slice\_01.png\textasciigrave{}, ..., \textasciigrave{}prostate\_00\_t2\_slice\_14.png\textasciigrave{}.
- Each PNG should preserve the slice appearance and orientation from the input MRI data.
- Keep \textasciigrave{}/app/input/*\textasciigrave{} unchanged.
- Keep final outputs under \textasciigrave{}/app/artifacts\textasciigrave{}.
\end{instructionbox}

\begin{instructionbox}[System \& Software Operations - Application \& Environment Config - 051-hide-left-dock]
Could you help me remove the dock on the left side of the screen? Please hide the dock on the left side, optionally using GIMP if needed, and then quit the application.
\end{instructionbox}

\begin{instructionbox}[System \& Software Operations - Application \& Environment Config - 052-vignette-filter-window]
Given the file \textasciigrave{}/app/dog\_with\_background.png\textasciigrave{}, could you help me open the Vignette filter window, optionally using GIMP if needed? Once the window is open, please quit the application.
\end{instructionbox}

\begin{instructionbox}[Multimedia \& Design - Image Editing - 053-make-palette-png]
Given the file \textasciigrave{}/app/computer.png\textasciigrave{}, could you help me set the image to be palette-based? You can do this optionally using GIMP if needed. Please keep the image visually the same while making the exported PNG palette-based, and save the result as \textasciigrave{}/app/palette\_computer.png\textasciigrave{}.
\end{instructionbox}

\begin{instructionbox}[Scientific \& Engineering - Medical Imaging - 054-prostate-label-obj]
Given \textasciigrave{}/app/input/case\_7f3a9c\_label.nii.gz\textasciigrave{} and \textasciigrave{}/app/input/task\_plan.json\textasciigrave{}, reconstruct a 3D prostate surface model with the installed 3D Slicer and write outputs under \textasciigrave{}/app/artifacts\textasciigrave{}.

Required outputs:
1. Read \textasciigrave{}/app/input/task\_plan.json\textasciigrave{}.
2. Read the provided label map. The nonzero voxels are the prostate segmentation.
3. Generate \textasciigrave{}/app/artifacts/prostate\_model.obj\textasciigrave{} as a real non-empty 3D prostate mesh derived from that label map.

Requirements:
- Use the real 3D Slicer installation in this environment. Headless use is fine.
- Keep \textasciigrave{}/app/input/*\textasciigrave{} completely unchanged.
- Keep all final outputs under \textasciigrave{}/app/artifacts\textasciigrave{}.
- Do not export the whole image volume, background, or placeholder geometry.
- The OBJ must be a valid mesh export of the prostate label.
\end{instructionbox}

\begin{instructionbox}[Multimedia \& Design - Image Editing - 055-resize-dog-layer]
Given the file \textasciigrave{}/app/dog\_with\_background\_two\_layers.xcf\textasciigrave{}, could you assist me with resizing the dog layer? I need to adjust its height to 512 pixels while maintaining the original aspect ratio. You can accomplish this optionally using GIMP if needed, and please save the final result as \textasciigrave{}/app/resized.png\textasciigrave{}.
\end{instructionbox}

\begin{instructionbox}[Multimedia \& Design - Image Editing - 056-move-textbox-left]
Given the file \textasciigrave{}/app/orange\_background.xcf\textasciigrave{}, could you help me shift the text box to the left side of the canvas? I keep accidentally selecting the image layer beneath it. You can do this optionally using GIMP if needed, and please save the result as \textasciigrave{}/app/leftside\_textbox.png\textasciigrave{}.
\end{instructionbox}

\begin{instructionbox}[System \& Software Operations - Application \& Environment Config - 057-set-undo-steps-100]
Could you help me set the minimum number of undo steps to 100, optionally using GIMP if needed, and then quit the application?
\end{instructionbox}

\begin{instructionbox}[System \& Software Operations - Application \& Environment Config - 058-autosave-3-minutes]
Given the file \textasciigrave{}/home/agent/.config/libreoffice/4/user/registrymodifications.xcu\textasciigrave{}, could you help me enable auto-save every 3 minutes so that I don't need to hit "ctrl-s" that much, optionally using LibreOffice Impress if needed? Please make sure to quit the application when finished so the saved preference is written to the file.
\end{instructionbox}

\begin{instructionbox}[Office \& Productivity - Presentations - 059-make-summary-slide]
Given the file \textasciigrave{}/app/Forests.pptx\textasciigrave{}, I am making a presentation for tomorrow and need to summarize the contents onto one slide. Could you help me create that using the built-in LibreOffice Impress "Summary Slide" feature? Please use that command's generated slide from the deck's slide titles, leave the generated slide at the end of the presentation, and do not manually write a prose summary. Please save the updated presentation in place at \textasciigrave{}/app/Forests.pptx\textasciigrave{}.
\end{instructionbox}

\begin{instructionbox}[Office \& Productivity - Presentations - 060-underline-red-text]
Given the file \textasciigrave{}/app/154\_3.pptx\textasciigrave{}, could you help me edit it, optionally using LibreOffice Impress if needed? Please underline the body of the slide only (without the title and table) in dark red 2, and change the font color of the whole slide (title, body, and table) to dark red 2. Finally, save the updated presentation in place at \textasciigrave{}/app/154\_3.pptx\textasciigrave{}.
\end{instructionbox}

\begin{instructionbox}[Office \& Productivity - Presentations - 061-insert-features-table]
Given the file \textasciigrave{}/app/41\_3.pptx\textasciigrave{}, could you help me insert a table with 5 rows and 2 columns into the "Features" slide, optionally using LibreOffice Impress if needed? Please save the updated presentation in place at \textasciigrave{}/app/41\_3.pptx\textasciigrave{}.
\end{instructionbox}

\begin{instructionbox}[Office \& Productivity - Presentations - 062-add-baseball-audio]
Given the file \textasciigrave{}/app/Mady\_and\_Mia\_Baseball.pptx\textasciigrave{}, I am making a presentation about the history of baseball. I want to add an introduction audio using the available file \textasciigrave{}/app/Baseball.mp3\textasciigrave{}, but I do not know how. Could you help me add this audio into my presentation file, optionally using LibreOffice Impress if needed? Please save the updated presentation in place at \textasciigrave{}/app/Mady\_and\_Mia\_Baseball.pptx\textasciigrave{}.
\end{instructionbox}

\begin{instructionbox}[Office \& Productivity - Presentations - 063-red-slide-numbers]
Given the file \textasciigrave{}/app/saa-format-guide.pptx\textasciigrave{}, I am preparing a presentation and the slide numbers are barely visible to me. Could you help me change the color of the slide numbers to red, optionally using LibreOffice Impress if needed? Please save the updated presentation in place at \textasciigrave{}/app/saa-format-guide.pptx\textasciigrave{}.
\end{instructionbox}

\begin{instructionbox}[Office \& Productivity - Presentations - 064-stretch-cover-image]
Given the file \textasciigrave{}/app/CPD\_Background\_Investigation\_Process.pptx\textasciigrave{}, optionally using LibreOffice Impress if needed, I want to turn the rectangular image of Columbus on the first page into a cover page. Could you help me stretch this image to fill the entire page, keeping its proportions and centering it? Please save the updated presentation in place at \textasciigrave{}/app/CPD\_Background\_Investigation\_Process.pptx\textasciigrave{}.
\end{instructionbox}

\begin{instructionbox}[Office \& Productivity - Presentations - 065-resize-slide3-slide6]
Given the file \textasciigrave{}/app/42\_2.pptx\textasciigrave{}, could you help me update it, optionally using LibreOffice Impress if needed? Please set the height of the picture on slide 3 to 20cm, and change the font size of all textboxes on slide 6 to 40pt. Finally, save the updated presentation in place at \textasciigrave{}/app/42\_2.pptx\textasciigrave{}.
\end{instructionbox}

\begin{instructionbox}[Office \& Productivity - Presentations - 066-set-slide-image-heights]
Given the file \textasciigrave{}/app/30\_1.pptx\textasciigrave{}, could you help me change the picture's height to 20cm, 30cm, and 25cm on slides 3, 4, and 6 respectively, optionally using LibreOffice Impress if needed? Please save the updated presentation in place at \textasciigrave{}/app/30\_1.pptx\textasciigrave{}.
\end{instructionbox}

\begin{instructionbox}[Office \& Productivity - Presentations - 067-align-slide-textboxes]
Given the file \textasciigrave{}/app/38\_1.pptx\textasciigrave{}, could you help me adjust some text alignments, optionally using LibreOffice Impress if needed? Please align the text of the first textbox on slide 3 to the right, on slide 4 to the center, and on slide 5 to the left. Ensure that the alignment is applied correctly to each respective slide, and then save the updated presentation in place at \textasciigrave{}/app/38\_1.pptx\textasciigrave{}.
\end{instructionbox}

\begin{instructionbox}[Office \& Productivity - Presentations - 068-strike-first-two-lines]
Given the file \textasciigrave{}/app/New\_Club\_Spring\_2018\_Training.pptx\textasciigrave{}, I am checking our soccer club's to-do list for the last semester and adding a strikethrough to the lines we have already accomplished. Could you help me add a strikethrough to the first and second lines, optionally using LibreOffice Impress if needed? Please save the updated presentation in place at \textasciigrave{}/app/New\_Club\_Spring\_2018\_Training.pptx\textasciigrave{}.
\end{instructionbox}

\begin{instructionbox}[Office \& Productivity - Documents - 069-linux-ls-tutorial]
Could you help me compose a Linux tutorial, optionally using LibreOffice Writer if needed? I'd like to display the results of running the "ls" command in /home/user. Please execute this command and save a screenshot of the terminal as 'ls.png' on the Desktop.
\end{instructionbox}

\begin{instructionbox}[Multimedia \& Design - Image Editing - 070-edit-friend-image]
Given the files at \textasciigrave{}/home/user/Desktop/\textasciigrave{}, could you help me assist a friend who asked for help editing an image? Please make the necessary modifications to the picture according to the instructions in the requirements document. Once finished, please save the edited picture as "pic.jpg" to the final output path: \textasciigrave{}/home/user/Desktop/pic.jpg\textasciigrave{}. Thank you!
\end{instructionbox}

\begin{instructionbox}[Office \& Productivity - Documents - 071-paste-image-docx]
Given the .xcf file on the Desktop, could you help me copy the image and paste it into a document, optionally using LibreOffice Writer if needed? Please save the document as 'image.docx' on the Desktop. Final output path: \textasciigrave{}/home/user/Desktop/image.docx\textasciigrave{}.
\end{instructionbox}

\begin{instructionbox}[System \& Software Operations - OS \& File Operations - 072-save-speedtest-results]
Could you help me test the quality of the network environment my laptop is currently in? Please measure my network situation, optionally using speedtest.net if needed. Copy the results (such as those from speedtest.net/results) and save them to \textasciigrave{}\textasciitilde{}/Test/Speed/results.txt\textasciigrave{} (if the directory does not exist, please create it). Each metric should occupy one line, with the metric name and its value separated by a single space. Final output path: \textasciigrave{}/home/user/Test/Speed/results.txt\textasciigrave{}.
\end{instructionbox}

\begin{instructionbox}[System \& Software Operations - OS \& File Operations - 073-force-quit-frozen-doc]
Given the file my\_document.odt, the application I'm working in---optionally using LibreOffice Writer if needed---seems to have frozen and I can't get it to close normally. Could you help me force quit the application from the command line? I'm on Ubuntu, and I don't want to restart my computer or lose any other work I have open.
\end{instructionbox}

\begin{instructionbox}[Office \& Productivity - Documents - 074-apa-references-review]
Given the file \textasciigrave{}/home/user/Desktop/students work/case study.docx\textasciigrave{}, could you please help me review it, optionally using Microsoft Word if needed? I'm particularly interested in ensuring that the references section at the end of the document adheres to the APA 7th edition formatting guidelines. Please make any necessary adjustments if the current formatting does not align with APA 7 standards or if there are any errors, and save the final version back to \textasciigrave{}/home/user/Desktop/students work/case study.docx\textasciigrave{}.
\end{instructionbox}

\begin{instructionbox}[Scientific \& Engineering - Bioimage Analysis - 075-nuclei-locations]
Use the provided microscopy images to locate the nuclei/cells.

CellProfiler is available in the environment. The nucleus-detection settings are summarized below. Follow these settings as closely as practical to detect the nucleus centers from the nuclear stain image.

Inputs:
- \textasciigrave{}/app/input/images/1-162hrh2ax2.tif\textasciigrave{}
- \textasciigrave{}/app/input/images/1-162hrhoe2.tif\textasciigrave{}

Relevant pipeline behavior:
- Treat both inputs as grayscale images.
- Use filename rules to identify the channels:
  - the file whose name contains \textasciigrave{}hoe\textasciigrave{} is the nuclear stain image; use this as the nucleus detection input.
  - the file whose name contains \textasciigrave{}h2ax\textasciigrave{} is the green foci image; it is not needed for nucleus center detection.
- Do not group image sets.
- Detect primary objects named \textasciigrave{}Nuclei\textasciigrave{} from the nuclear stain image with these settings:
  - typical object diameter: minimum \textasciigrave{}120\textasciigrave{} pixels, maximum \textasciigrave{}300\textasciigrave{} pixels
  - discard objects outside that diameter range
  - discard objects touching the image border
  - threshold strategy: global
  - thresholding method: Otsu
  - threshold correction factor: \textasciigrave{}1.0\textasciigrave{}
  - threshold smoothing scale: \textasciigrave{}1.3488\textasciigrave{}
  - threshold bounds: lower \textasciigrave{}0.0\textasciigrave{}, upper \textasciigrave{}1.0\textasciigrave{}
  - two-class thresholding; assign the middle-intensity class to foreground if applicable
  - distinguish clumped objects by shape
  - draw dividing lines between clumped objects by shape
  - smoothing filter size for declumping: \textasciigrave{}10\textasciigrave{}
  - suppress local maxima closer than \textasciigrave{}7.0\textasciigrave{} pixels
  - fill holes after thresholding and declumping
  - automatically calculate the smoothing filter and local-maxima distance for declumping
  - maximum object count: \textasciigrave{}500\textasciigrave{}
- Report each detected nucleus by its object center, equivalent to CellProfiler's \textasciigrave{}Location\_Center\_X\textasciigrave{} and \textasciigrave{}Location\_Center\_Y\textasciigrave{} measurements.

Required final output:
- \textasciigrave{}/app/artifacts/nuclei\_locations.csv\textasciigrave{}

Write a CSV file with exactly these columns:

\textasciigrave{}\textasciigrave{}\textasciigrave{}csv
x,y
\textasciigrave{}\textasciigrave{}\textasciigrave{}

Each data row should contain one nucleus/cell center in pixel coordinates. The output must be generated from the provided input images.
\end{instructionbox}

\begin{instructionbox}[Multimedia \& Design - Video \& Audio Editing - 076-make-src-gif-clip]
Given the file \textasciigrave{}/home/user/Desktop/src.mp4\textasciigrave{}, could you help me create a 5-second animated GIF clip beginning at 00:03, optionally using VLC and GIMP if needed? Please save the final output to \textasciigrave{}/home/user/Desktop/src\_clip.gif\textasciigrave{}.
\end{instructionbox}

\begin{instructionbox}[Web \& Information - Academic Lookup - 077-corresponding-scholar-url]
You are given the paper PDF \textasciigrave{}/app/shi17a.pdf\textasciigrave{}.

Use command-line tools to identify the Google Scholar profile URL of the paper's corresponding author.

Write exactly one URL to \textasciigrave{}/app/corresponding\_author\_scholar\_url.txt\textasciigrave{}.

Requirements:
1. The file must contain only the final \textasciigrave{}https://scholar.google.com/citations...\textasciigrave{} profile URL.
2. A trailing newline is fine, but do not add any other text.
3. Keep \textasciigrave{}/app/shi17a.pdf\textasciigrave{} unchanged.
\end{instructionbox}

\begin{instructionbox}[Office \& Productivity - Documents - 078-add-wednesday-lecture]
Given the spreadsheet \textasciigrave{}/home/user/Desktop/Course Timetable.xlsx\textasciigrave{}, could you help me add a two-hour lecture slot scheduled for every Wednesday at 12 PM? It seems I accidentally omitted that when setting up my schedule. I'd appreciate you taking care of that for me. Thanks!
\end{instructionbox}

\begin{instructionbox}[Multimedia \& Design - Video \& Audio Editing - 079-fix-mp3-metadata]
Given the MP3 files in \textasciigrave{}/home/user/Music/\textasciigrave{}, which have blank metadata but are already named with their artists and titles, could you help me fix the "title" and "artist" metadata, optionally using Picard or Kid3 if needed?
\end{instructionbox}

\begin{instructionbox}[Multimedia \& Design - Format Conversion - 080-convert-novel-epub]
The TXT chapters for the web novel are in \textasciigrave{}/home/user/Documents/Novels/Pass Through/\textasciigrave{}.

The EPUB conversion tool pages are already open in Chrome (Chromium in this environment) if you want to use them.

Convert the novel to EPUB for easy reading on mobile or Kindle, and save the result in the same directory using the novel title as the filename.

Accepted output filenames are \textasciigrave{}Pass Through.epub\textasciigrave{}, \textasciigrave{}Pass\_Through.epub\textasciigrave{}, or \textasciigrave{}pass\_through.epub\textasciigrave{}.
\end{instructionbox}

\begin{instructionbox}[Office \& Productivity - Presentations - 081-video-frame-slide-bg]
Extract the frame at \textasciigrave{}00:08\textasciigrave{} from the video, set it as the true background image of slide 2, and save the presentation in place.
Inputs:
- \textasciigrave{}/app/landscape.mp4\textasciigrave{}
- \textasciigrave{}/app/Robotic\_Workshop\_Infographics.pptx\textasciigrave{}
Output: \textasciigrave{}/app/Robotic\_Workshop\_Infographics.pptx\textasciigrave{}.
Use command-line tools and save only the durable output artifacts described above.
\end{instructionbox}

\begin{instructionbox}[Office \& Productivity - Presentations - 082-brighten-slide-image]
Given the file \{path\}, could you help me enhance the brightness of the image on the second slide, as it looks a bit too dim? You can optionally use PowerPoint if needed. Please save the adjusted image on the Desktop and name it "background.png". Thank you!
\end{instructionbox}

\begin{instructionbox}[Office \& Productivity - Documents - 083-gemini-responses-docx]
From \textasciigrave{}/app/llm\_answers.json\textasciigrave{}, Copy each Gemini response exactly as it appears in the JSON file's source text (i.e., the escaped form). Do NOT JSON-decode the strings. \textbackslash{}n stays as \textbackslash{}n, \textbackslash{}" stays as \textbackslash{}". The .docx should look identical to what you'd see opening the JSON in a text editor. 
Specifically look at the sentences in the responses that contain \textasciigrave{}Iliad\textasciigrave{}.
Put each response in its own paragraph with a blank line between responses, and highlight every occurrence of \textasciigrave{}Iliad\textasciigrave{} word.
Input: \textasciigrave{}/app/llm\_answers.json\textasciigrave{}.
Output: \textasciigrave{}/app/gemini\_results.docx\textasciigrave{}.
Use command-line tools and save only the durable output artifacts described above.
\end{instructionbox}

\begin{instructionbox}[Scientific \& Engineering - Engineering Simulation - 084-plate-thermal-sim]
Given \textasciigrave{}/app/input/simple\_plate\_regions.stl\textasciigrave{} and \textasciigrave{}/app/input/simple\_plate\_params.json\textasciigrave{}, create and run an OpenFOAM simulation for the simple heated plate.

Use the installed OpenFOAM tooling. Keep \textasciigrave{}/app/input/*\textasciigrave{} unchanged.

Simulation requirements:
- Use the geometry and named surface regions from \textasciigrave{}simple\_plate\_regions.stl\textasciigrave{}.
- Follow the physical parameters, boundary conditions, time controls, and numerics in \textasciigrave{}simple\_plate\_params.json\textasciigrave{}.
- Run the transient \textasciigrave{}laplacianFoam\textasciigrave{} case to \textasciigrave{}120 s\textasciigrave{}.
- Preserve the named patches: \textasciigrave{}heatSource\textasciigrave{}, \textasciigrave{}topRest\textasciigrave{}, \textasciigrave{}bottomSink\textasciigrave{}, \textasciigrave{}xmin\textasciigrave{}, \textasciigrave{}xmax\textasciigrave{}, \textasciigrave{}ymin\textasciigrave{}, and \textasciigrave{}ymax\textasciigrave{}.

Required outputs:
1. Create \textasciigrave{}/app/artifacts/simple\_plate\_case\textasciigrave{} containing the runnable OpenFOAM case and the generated final field at \textasciigrave{}/app/artifacts/simple\_plate\_case/120/T\textasciigrave{}.
2. Keep the solver log at \textasciigrave{}/app/artifacts/simple\_plate\_case/log.laplacianFoam\textasciigrave{}.
3. Write \textasciigrave{}/app/artifacts/simple\_plate\_result\_120s.json\textasciigrave{} as a nested JSON object, not with flattened dotted keys. Use this exact shape:

\textasciigrave{}\textasciigrave{}\textasciigrave{}json
\{
  "case\_name": "simple\_plate",
  "final\_time\_s": "\textless{}final simulation time in seconds\textgreater{}",
  "mesh": \{
    "cell\_count": "\textless{}integer\textgreater{}",
    "point\_count": "\textless{}integer\textgreater{}",
    "face\_count": "\textless{}integer\textgreater{}",
    "internal\_face\_count": "\textless{}integer\textgreater{}",
    "patch\_face\_counts": \{
      "heatSource": "\textless{}integer\textgreater{}",
      "topRest": "\textless{}integer\textgreater{}",
      "bottomSink": "\textless{}integer\textgreater{}",
      "xmin": "\textless{}integer\textgreater{}",
      "xmax": "\textless{}integer\textgreater{}",
      "ymin": "\textless{}integer\textgreater{}",
      "ymax": "\textless{}integer\textgreater{}"
    \}
  \},
  "metrics": \{
    "internal\_min\_temperature\_K": "\textless{}number\textgreater{}",
    "internal\_max\_temperature\_K": "\textless{}number\textgreater{}",
    "top\_surface\_min\_temperature\_K": "\textless{}number\textgreater{}",
    "top\_surface\_max\_temperature\_K": "\textless{}number\textgreater{}",
    "reported\_min\_temperature\_K": "\textless{}number\textgreater{}",
    "reported\_max\_temperature\_K": "\textless{}number\textgreater{}"
  \}
\}
\textasciigrave{}\textasciigrave{}\textasciigrave{}

Use numeric JSON values in your final file, not strings. The placeholders above only illustrate the required nesting and key names; compute all values from your actual run.
4. Write \textasciigrave{}/app/artifacts/simple\_plate\_top\_view.svg\textasciigrave{}, a top-view SVG visualization of the plate, heater patch, and simulated temperature result. This is for manual inspection in the artifact folder.

The verifier compares these values with hidden ground truth using 1\% relative tolerance.
\end{instructionbox}

\begin{instructionbox}[System \& Software Operations - Software Development - 085-webext-happy-scaffold]
Create the requested initial web-extension project scaffold with https://webext.eu and save the generated project tree under the output directory.
Output: \textasciigrave{}/app/happy-extension\textasciigrave{}.
Required scaffold semantics:
- project name: \textasciigrave{}happy-extension\textasciigrave{}
- version: \textasciigrave{}0.0.1\textasciigrave{}
- leave the description blank
- include a background script at \textasciigrave{}/app/happy-extension/background\_script.js\textasciigrave{}
- include a browser action popup under \textasciigrave{}/app/happy-extension/browserAction/\textasciigrave{}
- other extension features are not required
Produce the unzipped project tree directly under \textasciigrave{}/app/happy-extension\textasciigrave{}.
Use command-line tools and save only the durable output artifacts described above.
\end{instructionbox}

\begin{instructionbox}[Office \& Productivity - Documents - 086-futian-checkin-addresses]
Update the document with the required location list and save the document in place.
Input: \textasciigrave{}/app/AllLocations.docx\textasciigrave{}.
Output: \textasciigrave{}/app/AllLocations.docx\textasciigrave{}.
Write five addresses in Chinese for 24-hour self-service check-in machines in Futian District, Shenzhen. Keep the content focused on Futian District machine locations rather than a general travel note or addresses from other districts.
Use command-line tools and save only the durable output artifacts described above.
\end{instructionbox}

\begin{instructionbox}[Office \& Productivity - Documents - 087-spreadsheet-to-doc-table]
Transfer the data from the current sheet of \textasciigrave{}/app/OSP\_Envelope\_Price-List\_2023\_5000.xlsx\textasciigrave{} into a table in \textasciigrave{}/app/price.docx\textasciigrave{}, preserving the spreadsheet's original table formatting as closely as possible, and save the document.
Input: \textasciigrave{}/app/OSP\_Envelope\_Price-List\_2023\_5000.xlsx\textasciigrave{}.
Output: \textasciigrave{}/app/price.docx\textasciigrave{}.
Use command-line tools and save only the durable output artifacts described above.
\end{instructionbox}

\begin{instructionbox}[Multimedia \& Design - Image Editing - 088-extract-presenter-photos]
The event photos are already in \textasciigrave{}/home/user/Desktop/IDS LLM seminar/\textasciigrave{}, and your shell starts in \textasciigrave{}/home/user/Desktop\textasciigrave{}.

Please sift through the photos in \textasciigrave{}IDS LLM seminar\textasciigrave{} and extract the ones featuring the presenter Tao Yu into \textasciigrave{}/home/user/Desktop/presenter/\textasciigrave{}. Use Tao Yu's public profile or official web photo as the visual reference, and include every photo where he appears anywhere in the frame, including distant or group shots. Then create \textasciigrave{}/home/user/Desktop/presenter.zip\textasciigrave{} from that folder so that unzipping the archive recreates the \textasciigrave{}presenter/\textasciigrave{} folder on the Desktop. Keep both the extracted folder and the zip archive on the Desktop, not inside \textasciigrave{}IDS LLM seminar/\textasciigrave{}.
\end{instructionbox}

\begin{instructionbox}[System \& Software Operations - Application \& Environment Config - 089-install-recommended-exts]
Given the file \textasciigrave{}/home/user/Desktop/Recommended\_plugin\_list.docx\textasciigrave{}, my friend, who is a "plugin guru," recommended some good plug-ins to me. Could you help me go to the Chrome Web Store and install all the listed plug-ins into the default Google Chrome profile so they remain listed as installed extensions after Chrome restarts?
\end{instructionbox}

\begin{instructionbox}[Office \& Productivity - Spreadsheets - 090-ecs-grf-pass-rates]
Use the ECS and GRF PDF folders to update the supported-rate workbook by counting the ECS documents and organizing each school's pass rate by year in percentage form. Save the workbook in place.
Inputs:
- \textasciigrave{}/app/Fundings/ecs\textasciigrave{}
- \textasciigrave{}/app/Fundings/grf\textasciigrave{}
- \textasciigrave{}/app/Fundings/supported\_rate.xlsx\textasciigrave{}
Output: \textasciigrave{}/app/Fundings/supported\_rate.xlsx\textasciigrave{}.
Use command-line tools and save only the durable output artifacts described above.
\end{instructionbox}

\begin{instructionbox}[Office \& Productivity - Spreadsheets - 091-add-receipts-bookkeeping]
Given the workbook \textasciigrave{}/app/my\_bookkeeping.xlsx\textasciigrave{} and the receipt files \textasciigrave{}/app/receipt\_0.jpeg\textasciigrave{}, \textasciigrave{}/app/receipt\_1.jpg\textasciigrave{}, \textasciigrave{}/app/receipt\_2.jpg\textasciigrave{}, \textasciigrave{}/app/receipt\_3.pdf\textasciigrave{}, and \textasciigrave{}/app/receipt\_4.jpg\textasciigrave{}, use command-line tools to update the bookkeeping sheet with the recent transactions shown in those files. Save the completed spreadsheet in place at \textasciigrave{}/app/my\_bookkeeping.xlsx\textasciigrave{}.
\end{instructionbox}

\begin{instructionbox}[System \& Software Operations - Software Development - 092-fix-tetris-bug]
Given the file \{path\}, could you help me fix a bug in my code? I've recently been playing around with developing a small Python-based Tetris game, and while I've finished most of it, something goes wrong in certain cases. Specifically, when I press 'up' to rotate, the whole program crashes. Please fix the bugs in the code, optionally using Python to run it if needed.
\end{instructionbox}

\begin{instructionbox}[System \& Software Operations - OS \& File Operations - 093-merge-txt-document]
Given the files in your vscode project, could you help me merge the contents of all the .txt files into a single document, optionally using LibreOffice Writer if needed? No merging separator is required. Please ensure the overall font size of the document is set to 10, and save the final output to \textasciigrave{}/home/user/Desktop/concat.docx\textasciigrave{}.
\end{instructionbox}

\begin{instructionbox}[Web \& Information - Academic Lookup - 094-author-homepage-bookmarks]
The paper PDF is \textasciigrave{}/app/2206.08853.pdf\textasciigrave{}.

I'm really enjoying this paper. Could you please locate the personal webpages of the initial author and the last three authors?

Please include them in a Chrome browser bookmark folder titled \textasciigrave{}Liked Authors\textasciigrave{} under the \textasciigrave{}Bookmarks bar\textasciigrave{}.
\end{instructionbox}

\begin{instructionbox}[Web \& Information - Web Archiving - 095-save-apple-searching-page]
Given the website \textasciigrave{}https://developer.apple.com/design/human-interface-guidelines/searching\textasciigrave{}, could you help me obtain a local version of the blog's content to facilitate my own revisions? Please retain the primary content on the page, specifically from 'searching' to just before 'resources'. Optionally using Microsoft Word if needed, please save this blog content as "notes.docx" to the final output path: \textasciigrave{}/home/user/Desktop/notes.docx\textasciigrave{}.
\end{instructionbox}

\begin{instructionbox}[Office \& Productivity - Documents - 096-first-author-table]
Given the files in the folder, could you help me extract the name, e-mail, and affiliation of the first author from each paper? Please organize the extracted data into an Excel table---optionally using Excel if needed---and make sure to include headers for each field. Keep the workbook's default worksheet name unchanged. Afterward, sort the authors alphabetically by their full names and save the resulting table to \textasciigrave{}\textasciitilde{}/authors.xlsx\textasciigrave{} (final output path: \textasciigrave{}/home/user/authors.xlsx\textasciigrave{}).
\end{instructionbox}

\begin{instructionbox}[Office \& Productivity - Documents - 097-slides-to-document]
Given the file \{path\}, could you help me convert it into an editable document, optionally using Impress and Writer if needed? Simply extract the visible text from the presentation slides themselves, excluding speaker notes, comments, metadata, and any generated slide labels, and place it into \textasciigrave{}/home/user/Desktop/script.docx\textasciigrave{}; I'll handle the reformatting. Thank you!
\end{instructionbox}

\begin{instructionbox}[Office \& Productivity - Email \& Messaging - 098-extract-email-doc-image]
From the most recent email in the \textasciigrave{}Notes\textasciigrave{} folder, extract the first image from the attached DOC file and save that image to the output path as the desktop-background artifact.
Input: \textasciigrave{}/app/.thunderbird\textasciigrave{}.
Output: \textasciigrave{}/app/background.png\textasciigrave{}.
Use command-line tools and save only the durable output artifacts described above.
\end{instructionbox}

\begin{instructionbox}[Web \& Information - Academic Lookup - 099-professor-contact-info]
Given the file \textasciigrave{}/home/user/Desktop/Professor\_Contact.xlsx\textasciigrave{}, could you help me collect the contact information of the professors whose homepage links are listed inside? Please complete the form by adding their respective email addresses, optionally using Excel if needed. Once you are finished, please save the final output to \textasciigrave{}/home/user/Desktop/Professor\_Contact.xlsx\textasciigrave{}.
\end{instructionbox}

\begin{instructionbox}[Web \& Information - Public Reference - 100-name-mountain-photos]
Given the files in the Pictures directory, there are several pictures of mountains whose names I don't know. Could you help me identify these pictures and change their filenames to the names of the mountains shown in them? Use each mountain's standard common name rather than a descriptive phrase, and avoid adding redundant mountain terms to names that already include one. When renaming the files, please do not replace spaces with underscores.
\end{instructionbox}

\begin{instructionbox}[Multimedia \& Design - Image Editing - 101-rearrange-warm-tiles]
Split the image into three equal vertical sections, then rearrange them in order with a gradient of warm tones, progressively becoming warmer from left to right
Input: \textasciigrave{}/app/tilearray.png\textasciigrave{}.
Output: \textasciigrave{}/app/rearranged.png\textasciigrave{}.
Use command-line tools and save only the durable output artifacts described above.
\end{instructionbox}

\begin{instructionbox}[Office \& Productivity - Email \& Messaging - 102-daily-email-report]
Create \textasciigrave{}/app/report.xlsx\textasciigrave{} from the latest five emails in the \textasciigrave{}daily\textasciigrave{} folder, ordered from earliest to most recent, with columns \textasciigrave{}sender\_name\textasciigrave{}, \textasciigrave{}sender\_address\textasciigrave{}, \textasciigrave{}subject\textasciigrave{}, \textasciigrave{}CC\textasciigrave{}, and \textasciigrave{}number\_of\_attachments\textasciigrave{}.
Input: \textasciigrave{}/app/.thunderbird\textasciigrave{}.
Output: \textasciigrave{}/app/report.xlsx\textasciigrave{}.
Use command-line tools and save only the durable output artifacts described above.
\end{instructionbox}

\begin{instructionbox}[Web \& Information - Web Archiving - 103-download-blog-pdfs]
I browsed a lot of interesting blog articles today. I hope to store these articles in my local designated folder just like Zotero stores papers. Please download the blogs opening now in pdf format and save them in their title to \textasciigrave{}/home/user/Documents/Blog\textasciigrave{}.
\end{instructionbox}

\begin{instructionbox}[Office \& Productivity - Spreadsheets - 104-arxiv-llm-workbook]
Update the workbook with the paper list for new foundation language models issued via arXiv Daily on 2023-10-11, including the paper titles, arXiv PDF links, and dates required by the sheet. Save the workbook in place.
Input: \textasciigrave{}/app/New Large Language Models.xlsx\textasciigrave{}.
Output: \textasciigrave{}/app/New Large Language Models.xlsx\textasciigrave{}.
Use command-line tools and save only the durable output artifacts described above.
\end{instructionbox}

\begin{instructionbox}[Web \& Information - Public Reference - 105-search-cell-b6]
Given the file \{path\}, could you help me copy the data from Cell B6 and search for it in a web browser? You can optionally use LibreOffice Calc and Google Chrome if needed. Open the relevant page and, when finished, leave that exact page selected as the active tab in the foreground window.
\end{instructionbox}

\begin{instructionbox}[System \& Software Operations - OS \& File Operations - 106-create-charles-ssh-user]
Create a normal shell-login user named \textasciigrave{}charles\textasciigrave{} with password \textasciigrave{}Ex@mpleP@55w0rd!\textasciigrave{}.

Set \textasciigrave{}charles\textasciigrave{}'s home directory to \textasciigrave{}/home/test1\textasciigrave{} and make that directory writable by \textasciigrave{}charles\textasciigrave{}. Do not make the account SFTP-only or assign a non-login shell.
\end{instructionbox}

\begin{instructionbox}[Office \& Productivity - Email \& Messaging - 107-attach-aws-bill]
Attach the my AWS bill to the email. The bill is stored at \textasciigrave{}\textasciitilde{}/aws-bill.pdf\textasciigrave{}. Don't close it or send it. I haven't finish all the contents.
\end{instructionbox}

\begin{instructionbox}[Office \& Productivity - Email \& Messaging - 108-create-mail-folders]
Could you help me create two local folders named \textasciigrave{}COMPANY\textasciigrave{} and \textasciigrave{}UNIVERSITY\textasciigrave{}, optionally using Thunderbird if needed? Please quit the application when you are done so the folders are saved.
\end{instructionbox}

\begin{instructionbox}[Office \& Productivity - Email \& Messaging - 109-setup-outlook-account]
Help me access my outlook account with address \textasciigrave{}anonym-x2024@outlook.com\textasciigrave{} and password \textasciigrave{}password\textasciigrave{} in Thunderbird. Just fill in the information and stay on that page. I will check it manually later.
\end{instructionbox}

\begin{instructionbox}[Multimedia \& Design - Video \& Audio Editing - 110-rotate-macintosh-video]
Given the file \textasciigrave{}\textasciitilde{}/Desktop/flipped\_1984\_Apple\_Macintosh\_Commercial.mp4\textasciigrave{}, could you help me turn this video the right way up? Once it's flipped around, please save the final output file exactly at \textasciigrave{}\textasciitilde{}/1984\_Apple\_Macintosh\_Commercial.mp4\textasciigrave{}. You can optionally use VLC Media Player (a headless \textasciigrave{}vlc\textasciigrave{} or \textasciigrave{}cvlc\textasciigrave{} workflow is acceptable in this terminal-first Harbor environment) if needed. Only the final file at that exact path counts for the score.
\end{instructionbox}

\begin{instructionbox}[Multimedia \& Design - Video \& Audio Editing - 111-set-video-wallpaper]
Given the file \textasciigrave{}\textasciitilde{}/Desktop/Interstellar Movie - Official Trailer.mp4\textasciigrave{}, could you help me make a part of the video my computer's background picture? You can optionally use VLC Media Player (a headless \textasciigrave{}vlc\textasciigrave{} or \textasciigrave{}cvlc\textasciigrave{} workflow is acceptable in this terminal-first Harbor environment) if needed. Please leave the final wallpaper set when you finish, as no extra output file is required.
\end{instructionbox}

\begin{instructionbox}[Multimedia \& Design - Video \& Audio Editing - 112-capture-video-frame]
Given the file \textasciigrave{}\textasciitilde{}/Desktop/Interstellar Movie - Official Trailer.mp4\textasciigrave{}, could you help me snap a photo of a scene from the video? You can optionally use VLC Media Player if needed, and a headless \textasciigrave{}vlc\textasciigrave{} or \textasciigrave{}cvlc\textasciigrave{} workflow is perfectly acceptable in this terminal-first Harbor environment. Please save the final output file exactly at \textasciigrave{}\textasciitilde{}/Desktop/interstellar.png\textasciigrave{}, as only the file at that exact path will count for the score.
\end{instructionbox}

\begin{instructionbox}[System \& Software Operations - Application \& Environment Config - 113-add-folders-workspace]
Given the file \textasciigrave{}/app/project.code-workspace\textasciigrave{}, could you help me add the folders \textasciigrave{}/app/data1\textasciigrave{} and \textasciigrave{}/app/data2\textasciigrave{} to this workspace, optionally using VS Code if needed?
\end{instructionbox}

\begin{instructionbox}[Scientific \& Engineering - Bioimage Analysis - 114-nuclei-csv-open]
Use the provided CellProfiler pipeline and microscopy images to reproduce the hidden nuclei measurement table.

Visible inputs:
- \textasciigrave{}/app/input/config/ExampleSpeckles.cppipe\textasciigrave{}
- \textasciigrave{}/app/input/images/1-162hrh2ax2.tif\textasciigrave{}
- \textasciigrave{}/app/input/images/1-162hrhoe2.tif\textasciigrave{}

Required final output:
- \textasciigrave{}/app/artifacts/Nuclei.csv\textasciigrave{}

The output must be generated from the provided inputs. Do not hand-edit or fabricate the CSV.
\end{instructionbox}

\begin{instructionbox}[System \& Software Operations - Application \& Environment Config - 115-remove-explorer-find-key]
Given the file \textasciigrave{}\textasciitilde{}/.config/Code/User/keybindings.json\textasciigrave{}, could you help me remove the shortcut \textasciigrave{}ctrl+f\textasciigrave{} for Tree View Find (Explorer search) in the VS Code Explorer view to avoid a shortcut conflict, optionally using VS Code if needed?
\end{instructionbox}

\begin{instructionbox}[Office \& Productivity - Documents - 116-bottom-left-page-numbers]
Given the file \textasciigrave{}/app/LibreOffice\_Open\_Source\_Word\_Processing.docx\textasciigrave{}, could you help me add page numbers to every page at the bottom left? You can optionally use LibreOffice Writer if needed. Please save the updated document in place at \textasciigrave{}/app/LibreOffice\_Open\_Source\_Word\_Processing.docx\textasciigrave{} and keep the rest of the document content unchanged.
\end{instructionbox}

\begin{instructionbox}[Office \& Productivity - Documents - 117-comma-text-to-table]
Given the file \textasciigrave{}/app/Graphemes\_Sound\_Letter\_Patterns.docx\textasciigrave{}, could you help me convert the comma-separated text in the document into a table? You can do this optionally using LibreOffice Writer if needed. Please keep the rest of the document content unchanged and save the updated document in place at \textasciigrave{}/app/Graphemes\_Sound\_Letter\_Patterns.docx\textasciigrave{}.
\end{instructionbox}

\begin{instructionbox}[Office \& Productivity - Documents - 118-set-times-new-roman]
Given the file \textasciigrave{}/app/Dublin\_Zoo\_Intro.docx\textasciigrave{}, could you help me change the font to \textasciigrave{}Times New Roman\textasciigrave{} throughout the text? You can optionally use LibreOffice Writer if needed. Please save the updated document in place at \textasciigrave{}/app/Dublin\_Zoo\_Intro.docx\textasciigrave{} and keep the rest of the document content unchanged.
\end{instructionbox}

\begin{instructionbox}[Office \& Productivity - Documents - 119-tabstop-sentence-split]
Given the file \textasciigrave{}/app/04 CHIN9505 EBook Purchasing info 2021 Jan.docx\textasciigrave{}, could you help me format it? You can optionally use LibreOffice Writer if needed. For each sentence, please keep the first three words left-aligned and move the remaining words to the right using tabstops, creating an empty space in the middle for photos. Finally, save the updated document in place at \textasciigrave{}/app/04 CHIN9505 EBook Purchasing info 2021 Jan.docx\textasciigrave{}, keeping the rest of the document content unchanged.
\end{instructionbox}

\end{document}